\documentclass[journal=jctcce,manuscript=article]{achemso}

\usepackage[version=3]{mhchem} 
\usepackage{amsfonts}
\usepackage{bm}
\usepackage{xcolor}
\usepackage{caption}
\usepackage{subcaption}
\usepackage{graphicx}
\usepackage{placeins}
\usepackage{hyperref}
\author{Stephan Thaler}
\email{stephan.thaler@tum.de}
\author{Gregor Doehner}
\author{Julija Zavadlav}
\email{julija.zavadlav@tum.de}
\affiliation{Professorship of Multiscale Modeling of Fluid Materials, Department of Engineering Physics and Computation,
TUM School of Engineering and Design, Technical University of Munich, Germany}
\alsoaffiliation{Munich Data Science Institute, Technical University of Munich, Germany}

\title[]
    {Scalable Bayesian Uncertainty Quantification for Neural Network Potentials: Promise and Pitfalls}

\abbreviations{}

\begin{document}
\begin{abstract}
Neural network (NN) potentials promise highly accurate molecular dynamics (MD) simulations within the computational complexity of classical MD force fields.
However, when applied outside their training domain, NN potential predictions can be inaccurate, increasing the need for Uncertainty Quantification (UQ).
Bayesian modeling provides the mathematical framework for UQ, but classical Bayesian methods based on Markov chain Monte Carlo (MCMC) are computationally intractable for NN potentials.
By training graph NN potentials for coarse-grained systems of liquid water and alanine dipeptide, we demonstrate here that scalable Bayesian UQ via stochastic gradient MCMC (SG-MCMC) yields reliable uncertainty estimates for MD observables.
We show that cold posteriors can reduce the required training data size and that for reliable UQ, multiple Markov chains are needed.
Additionally, we find that SG-MCMC and the Deep Ensemble method achieve comparable results, despite shorter training and less hyperparameter tuning of the latter.
We show that both methods can capture aleatoric and epistemic uncertainty reliably, but not systematic uncertainty, which needs to be minimized by adequate modeling to obtain accurate credible intervals for MD observables.
Our results represent a step towards accurate UQ that is of vital importance for trustworthy NN potential-based MD simulations required for decision-making in practice.
\end{abstract}

\section{Introduction}
Molecular dynamics (MD) simulations are the computational tool of choice to describe complex molecular phenomena. Their computational effort and accuracy depend on the
chosen potential energy model. Neural network (NN) potentials \cite{Behler2007, Schutt2017, Gilmer2017, Klicpera2020, Klicpera2020b, Ko2021, Batzner2022}, which model many-body interactions \cite{Noe2020, Schutt2017a}, promise MD simulations at ab initio accuracy \cite{Faber2017, Klicpera2020} within the computational complexity of classical molecular mechanics force fields.

The quality of NN potentials is limited by the scarcity of suitable training data \cite{Stocker2022}, given that data generation via computational quantum mechanics simulations and/or experiments is resource intensive. 
Hence, potentials are commonly applied outside their training domain due to the high-dimensional chemical space.
As NN potentials are data-driven black box models, predictions outside the training domain may be inaccurate or even unphysical \cite{Wang2019, Thaler_2021, Van2022}. 
This may hinder more widespread adoption of NN potentials in practical applications where less powerful but physically more constrained models are preferred \cite{Gal2022}.

Uncertainty quantification (UQ) can provide a remedy as it enables practitioners to quantify the trustworthiness of MD simulation predictions\cite{Angelikopoulos2012, Zavadlav2019,Wan2021}. 
Additionally, the availability of a UQ metric enables more efficient training data generation via active learning \cite{Smith2018, Zhang2019, Loeffler2020, Smith2021, Xie2021, Van2022}, as well as an adaptive combination of NN potentials with established classical force fields \cite{Imbalzano2021}. Bayesian statistics provides a mathematically rigorous approach to UQ. However, classical Bayesian inference schemes based on Markov Chain Monte Carlo (MCMC), such as Hamiltonian (or hybrid \cite{Duane1987}) Monte Carlo (HMC) \cite{Neal2011}, require an evaluation of the likelihood over the whole data set for each parameter update. Frequent full likelihood evaluations are prohibitively expensive for computationally demanding NNs and large data sets\cite{Kahle2022}.
Stochastic gradient MCMC (SG-MCMC) schemes \cite{Welling2011, Chen2014, Li2016, Nemeth2021, Lamb2020} enable scalable Bayesian UQ of NNs by computing stochastic estimates of the gradient of the likelihood on a mini-batch of data.
Stochastic variational inference \cite{Graves2011, Hoffman2013} represents another scalable Bayesian UQ method, while the Deep Ensemble \cite{Hansen90, Lakshminarayanan2017} method is a popular non-Bayesian \cite{Lakshminarayanan2017, Ovadia2019, Wenzel2020, Gal2022} alternative.

In the context of NN potentials, the Deep Ensemble method is in fact the most common UQ scheme \cite{Klicpera2020b, Imbalzano2021, Kahle2022}, but Dropout Monte Carlo \cite{Wen2020} and last-layer Gaussian Mixture Models \cite{Zhu2022} have also been applied.
In view of the poor performance of the Deep Ensemble method in an active learning context, \citeauthor{Kahle2022} \cite{Kahle2022} recently hypothesized that Bayesian approaches may provide more reliable uncertainty estimates for NN potentials. However, a comprehensive assessment of Bayesian UQ in the context of NN potentials is still outstanding. 

In this work, we investigate scalable Bayesian UQ of MD observables for simulations utilizing NN molecular models.
To this end, we first compare the UQ quality of a SG-MCMC method to the popular Deep Ensemble method and a gold-standard \cite{Gustafsson2020, Lamb2020, Nemeth2021, Gal2022} HMC sampler based on a Lennard Jones (LJ) system with a 2-body toy NN potential.
We then extend the comparison by learning graph NN potentials for coarse-grained (CG) systems of water and alanine dipeptide, demonstrating the practical applicability of SG-MCMC methods to fully-Bayesian modeling of graph NN potentials.
Additionally, we investigate the influence of so-called cold posteriors \cite{Wenzel2020} and the number of MCMC chains on the quality of Bayesian UQ. 
Finally, we advocate distinguishing between different sources of uncertainty; in particular, we highlight the importance of minimizing systematic uncertainties to obtain reliable credible intervals of MD observables.

\section{Methods}
In the following, we briefly summarize the employed SG-MCMC sampler as well as the Deep Ensemble method and continue with an outline of the Bayesian molecular modeling problem considered in this work.

\subsection{Sources of uncertainty}
The uncertainty in physical modeling can be divided into aleatoric, epistemic and systematic uncertainty \cite{Gal2022}. Aleatoric uncertainty refers to the inherent stochastic nature of the modeled process, which can be interpreted as randomness in the labels $\mathbf{y}$ for a given input $\mathbf{x}$  \cite{Gustafsson2020, Hullermeier2021}. Epistemic uncertainty refers to the uncertainty about the true hypothesis (model) within the considered hypothesis space (model family).
In contrast to aleatoric uncertainty, epistemic uncertainty can be reduced by gathering more data.
Finally, systematic uncertainty is caused by model misspecification, i.e., when the true data-generating process is not contained within the hypothesis space. Systematic uncertainty manifests itself in an inconsistency between the data and the hypothesis space \cite{Hullermeier2021}.

\subsection{Bayesian Modeling}
A probabilistic model $p(\mathbf{y}|\mathbf{x},\mathbf{\bm \theta})$ predicts the distribution of $\mathbf{y}$ reflecting the aleatoric uncertainty, given a training data set $\mathcal{D}=\{\mathbf{x}_i, \mathbf{y}_i \}_{i=1}^N$ of size $N$.
Bayesian UQ additionally aims to quantify the epistemic uncertainty resulting from the model fit to a finite amount of data \cite{Ovadia2019, Gal2022}. Instead of selecting a single set of model parameters $\bm \theta$, the Bayesian approach promises more robust predictions by marginalizing over $\bm \theta$ \cite{Wilson2020}. Hence, the goal of Bayesian UQ is to compute the posterior predictive distribution
\begin{equation}
    p(\mathbf{y}|\mathbf{x}, \mathcal{D}) = \int p(\mathbf{y}|\mathbf{x},\bm{\theta}) p(\bm{\theta}|\mathcal{D}) d\,\bm{\theta} \ ,
    \label{eq:posterior_predictive}
\end{equation}
where $p(\bm{\theta}|\mathcal{D})$ is the posterior distribution. The integral in eq.~\eqref{eq:posterior_predictive} is typically approximated by the Monte Carlo method:
\begin{equation}
    p(\mathbf{y}|\mathbf{x},\mathcal{D}) \approx  \frac{1}{N} \sum_{n=1}^N p(\mathbf{y}|\mathbf{x},\bm \theta_{n}) \ ,
    \label{eq:Monte_Carlo}
\end{equation}
where $\bm{\theta}_{n}$ represents the $n^\mathrm{th}$ model parameter set drawn from the posterior. 
Evaluating eq.~\eqref{eq:Monte_Carlo} requires sampling from the posterior distribution 
\begin{equation}
    p(\bm \theta|\mathcal{D}) = \frac{p(\mathcal{D}|\bm \theta)p(\bm \theta)}{p(\mathcal{D})} \propto \text{exp} \left( \frac{-\mathcal{U}(\bm{\theta})}{\mathcal{T}} \right) \ ,
    \label{eq:posterior_energy_relation}
\end{equation}
with likelihood $p(\mathcal{D}|\bm \theta)$ and prior $p(\bm{\theta})$.
In analogy to statistical mechanics, the posterior can be re-written to allow sampling from a Boltzmann-type distribution \cite{Neal2011}, with posterior potential energy
\begin{equation}
    \mathcal{U}(\bm{\theta}) = - \sum_{i=1}^{N}\text{log}\;p(\mathbf{y}_i|\mathbf{x}_i,\bm{\theta})-\text{log}\;p(\bm{\theta}) \ ,
    \label{eq:potential_energy_function}
\end{equation}
and posterior temperature $\mathcal{T}$, which is introduced as an additional hyperparameter. $\mathcal{T} = 1$ corresponds to the Bayesian posterior, while $\mathcal{T} < 1$ are sharper \cite{Izmailov2021} cold posteriors \cite{Wenzel2020}.

The gold-standard HMC \cite{Duane1987, Neal2011} method leverages the gradient $\nabla_\theta \mathcal{U}(\bm{\theta})$ to simulate Hamiltonian dynamics to generate parameter proposals for the Metropolis Hastings \cite{Hastings1970} (MH) acceptance step, which guarantees that the equilibrium distribution of the Markov chain corresponds to $p(\bm \theta|\mathcal{D})$.
The computation of $\nabla_{\bm \theta} \mathcal{U}(\bm{\theta})$  \cite{Neal2011} requires an evaluation of the (NN) model for the whole training data set $\mathcal{D}$ (eq.~\eqref{eq:potential_energy_function}), rendering HMC computationally intractable for training NN potentials \cite{Li2011, Kahle2022}.

\subsection{Stochastic gradient MCMC}
Stochastic gradient MCMC (SG-MCMC) methods \cite{Welling2011, Chen2014, Li2016, Nemeth2021} achieve enormous computational speed-ups by replacing $\nabla_{\bm{\theta}} \mathcal{U}(\bm{\theta})$ by a stochastic estimate over a mini-batch of data
\begin{equation}
    \nabla_{\bm{\theta}} \tilde{\mathcal{U}}(\bm{\theta}) = -\frac{N}{B}\sum_{i=1}^B \nabla_{\bm \theta} \text{log}\;p(\mathbf{y}_i|\mathbf{x}_i,\bm{\theta})- \nabla_{\bm{\theta}}\text{log}\;p(\bm{\theta}) \ ,
    \label{eq:mini_batch_gradient}
\end{equation}
where B is the mini-batch size.

The simplest SG-MCMC scheme is the Stochastic Gradient Langevin Dynamics (SGLD) method \cite{Welling2011}, which updates parameters according to
\begin{equation}
    \bm \theta_{k+1} = \bm \theta_{k} - \frac{\lambda_k}{2} \nabla_{\bm \theta} \tilde{\mathcal{U}}(\bm{\theta}) + \bm \eta_t \ ; \quad \bm \eta_t \sim \mathcal{N}(\mathbf{0}, \lambda_k \mathbf{I}) \ .
\end{equation}
The learning rate $\lambda_k$ is decreasing as a function of update step $k$ and $\bm \eta_t$ is a learning rate-dependent Gaussian noise vector. $\lambda_k$ typically follows a polynomial schedule \cite{Welling2011, Teh2016}:
\begin{equation}
    \lambda_k = a (k+1)^{-\gamma} \ ,
    \label{eq:lr_schedule}
\end{equation}
where $a$ is the initial learning rate and $\gamma$ is the decay rate.
To reduce the bias due to the omitted MH acceptance step, it is necessary to sample only below a certain learning rate threshold, given that the acceptance probability asymptotically converges to 1 for $\lambda \rightarrow 0$.
Hence, SGLD smoothly transitions from stochastic posterior maximization to asymptotically unbiased sampling from $p(\bm \theta|\mathcal{D})$ during training \cite{Welling2011, Teh2016}.
In our experiments, we employ a preconditioned version of SGLD (pSGLD) \cite{Li2016}, which uses a RMSProp \cite{Tieleman2012} preconditioner to simplify sampling the highly non-convex posterior of NNs \cite{Dauphin2014, Li2016}, as implemented in jax-sgmc \cite{Thaler2022c}.

\subsection{Deep Ensemble method}
Analogous to the Monte Carlo approximation in Bayesian UQ (eq.~\eqref{eq:Monte_Carlo}), the Deep Ensemble method \cite{Hansen90, Lakshminarayanan2017} estimates epistemic uncertainty from the statistics of predictions from an ensemble of NNs. However, instead of sampling models from the posterior, the ensemble of NNs is generated by minimizing a loss function via stochastic gradient descent, starting from different random NN weight initializations.
If desired, aleatoric uncertainty can be quantified by additionally predicting standard deviations and minimizing a negative log-likelihood loss \cite{Lakshminarayanan2017}.
While most authors consider the Deep Ensemble method non-Bayesian \cite{Lakshminarayanan2017, Ovadia2019, Wenzel2020, Gal2022}, \citeauthor{Wilson2020}\cite{Wilson2020} compellingly argue that it can also be interpreted as Bayesian model averaging.

\subsection{Neural Network Posterior Landscape}
The posterior distribution of NNs is high-dimensional, non-convex and multi-modal \cite{Dauphin2014, Fort2019, Wilson2020, Izmailov2021}.
The NNs of the Deep Ensemble typically converge to different posterior modes due to the strong decorrelation effect of different random weight initializations \cite{Fort2019, Wilson2020}.
Hence, the Deep Ensemble method performs a Bayesian model average of NNs corresponding to different approximate maximum a-posteriori (MAP) points on the NN posterior (assuming regularization terms that mimic the prior) \cite{Wilson2020}. 
The Deep Ensemble method therefore exploits the NN posterior multi-modality to estimate the uncertainty.
By contrast, most scalable Bayesian methods, including single-chain SG-MCMC and stochastic variational inference, have been found to typically approximate a single posterior mode only \cite{Ovadia2019, Gustafsson2020, Wilson2020}.
However, sampling multiple posterior modes is essential for robust UQ \cite{Wilson2020}.

\subsection{Multi-chain SG-MCMC}

Sampling the posterior with multiple randomly initialized SG-MCMC chains appears to be a promising approach. It combines Bayesian posterior exploration along the Markov chain with strong decorrelation from different random initializations, the benefits of which have been shown to be complementary \cite{Fort2019, Izmailov2021}.
Multi-chain SG-MCMC can be interpreted as a custom trade-off between the number of approximated posterior modes and the amount of Bayesian exploration per mode, with single-mode Bayesian methods and the Deep Ensemble method representing the two extreme cases. 

The computational training cost of the Deep Ensemble method and SG-MCMC can be estimated as $C * n_\mathrm{steps} * n_\mathrm{chains}$, where $n_\mathrm{chains}$ is the number of ensemble members (chains), $n_\mathrm{steps}$ is the number of parameter updates per ensemble member (chain) and $C$ is the cost per update. Training the different ensemble members (chains) can be parallelized trivially, if desired.

\subsection{Probabilistic Molecular Modeling}

\subsubsection{Maximum Likelihood Molecular Modeling}
The most common training scheme for atomistic (AT) NN potentials is to match the potential energy (possibly also forces and virial) of an underlying high-fidelity
model, usually a computational quantum mechanics scheme \cite{Smith2020},
given a training data set of $N_\mathrm{box}$ molecular states \cite{Noe2020}. This can be achieved by minimizing the mean squared error loss function
\begin{equation}
\label{eq:energy_matching_loss}
    L(\bm \theta) = \frac{1}{N_\mathrm{box}}\sum_{i=1}^{N_\mathrm{box}} [U_i - U_{i,\bm \theta} ]^2 \ ,
\end{equation}
where $U_i$ and $U_{i,\bm \theta}$ are the target and predicted potential energies of molecular state $i$, respectively. The predicted potential energy $U_{\bm \theta}(\mathbf{r})$ depends on atom positions $\mathbf{r}$.

Similarly, for CG systems, the NN potential can be trained via force matching (FM) \cite{Izvekov2005, Noid2008, Noid2008b, Wang2009}, i.e., matching the instantaneous force components $F_j$ acting on each CG particle as computed from the AT force field:
\begin{equation}
\label{eq:CG_FM_loss}
    L(\bm \theta) = \frac{1}{N_\mathrm{F}}\sum_{j=1}^{N_\mathrm{F}} [F_j - F_{j, \bm \theta}]^2 \ ,
\end{equation}
where $N_\mathrm{F}$ is 3 times the number of CG particles in the training data set. The predicted force components are computed from the CG NN potential $\mathbf{F}_{\bm \theta} = - \nabla_\mathbf{R} U^\mathrm{CG}_{\bm \theta}(\mathbf{R})$, which acts on CG coordinates $\mathbf{R} = \mathbf{M}(\mathbf{r})$. $\mathbf{M}$ is a linear function that maps from AT to CG coordinates.
For infinite data and model capacity, $U^\mathrm{CG}_{\bm \theta}(\mathbf{R})$ converges to the potential of mean force (PMF).
Given that multiple AT configurations map to the same CG configuration, there exists a lower bound of the loss in eq.~\eqref{eq:CG_FM_loss}, which corresponds to the loss of the PMF \cite{Noid2008, Wang2019}.

\subsubsection{Bayesian Molecular Modeling}
Assuming independent Gaussian homoscedastic aleatoric uncertainty with variance $\sigma^2_\mathrm{H}$, the 
probabilistic model of the energy matching task is $p(U|\mathbf{r},\bm{\theta}) \sim \mathcal{N}(U_{\bm \theta}(\mathbf{r}), \sigma^2_\mathrm{H})$.
In this case, the likelihood can be written similar to eq.~\eqref{eq:energy_matching_loss} as \cite{Kahle2022}
\begin{equation}
\begin{split}
    \label{eq:Likelihood_MD}
    p(\mathcal{D} | \bm \theta) &= \prod_{i=1}^{N_\mathrm{box}} \frac{1}{\sqrt{2 \pi \sigma^2_\mathrm{H}}} \exp{\left(-\frac{[U_i - U_{i,\bm \theta} ]^2}{2 \sigma^2_\mathrm{H}} \right)}\\ &= \left(\frac{1}{\sqrt{2 \pi \sigma^2_\mathrm{H}}}\right)^{N_\mathrm{box}} \exp{\left( -\frac{\sum_{i=1}^{N_\mathrm{box}}[U_i - U_{i,\bm \theta} ]^2}{2 \sigma^2_\mathrm{H}}  \right)} \ .
\end{split}
\end{equation}
The probabilistic model and the likelihood for the FM task follow analogously. The loss minima in eq.~\eqref{eq:energy_matching_loss} and \eqref{eq:CG_FM_loss} correspond to the likelihood maxima in eq.~\eqref{eq:Likelihood_MD}.

The aleatoric uncertainty is uncertainty inherent to the data. When learning atomistic models from simulation data, the aleatoric uncertainty stems from the data-generating simulation and is typically small. For CG systems, the non-injective CG mapping contributes significantly to the aleatoric uncertainty.
The variance of the aleatoric uncertainty $\sigma^2_\mathrm{H}$ is typically unknown a priori and we model it as a learnable parameter. Thus, the prior $p(\bm \theta) = p(\mathbf{w}) p(\sigma_\mathrm{H})$ is the product of a prior for the NN potential weights and biases $\mathbf{w}$ and a prior for the aleatoric uncertainty scale.

\subsection{Neural Network Potential}
We choose a graph NN potential, which is a state-of-the-art NN architecture that learns to extract predictive features from the molecular configuration in an end-to-end manner instead of relying on hand-crafted descriptors \cite{Schutt2017, Gilmer2017}. Specifically, we select
our previously published implementation \cite{Thaler_2021} of the DimeNet++ \cite{Klicpera2020, Klicpera2020b} potential.
We set all hyperparameters to their default values, including the graph cut-off radius of $r_\mathrm{cut} = 0.5$ nm, except for embedding sizes, which we reduce by factor 4 for computational speed-up.
We select a Gaussian prior over all learnable weights and biases $p(\mathbf{w}) \sim \mathcal{N}(\mathbf{0}, 10^2\mathbf{I})$, except for the radial Bessel frequencies \cite{Klicpera2020}, which we model by a uniform distribution.

Given that DimeNet++ trained via FM tends to yield unstable MD simulations \cite{Fu2022}, we augment the NN potential with a fixed, physics-informed "prior" potential $U^\mathrm{prior}(\mathbf{R})$ \cite{Das2009, Wang2019, Husic2020}:
\begin{equation}
\label{eq:prior_addition}
    U^\mathrm{CG}_{\bm \theta}(\mathbf{R}) = U_{\bm \theta}^{\mathrm{NN}}(\mathbf{R}) + U^\mathrm{prior}(\mathbf{R}) \ .
\end{equation}
Note that $U^\mathrm{prior}(\mathbf{R})$ is not a prior in the Bayesian sense, but rather a physics-informed initialization that enforces physically reasonable predictions in phase-space regions unconstrained by the training data \cite{Wang2019, Thaler_2021, Thaler2022} (see supplementary methods~1 for more details).

\section{Results}
\begin{figure}[htbp]
    \centering
    \begin{subfigure}[b]{0.32\textwidth}   
        \centering 
        \raisebox{8mm}{\includegraphics[width=\textwidth]{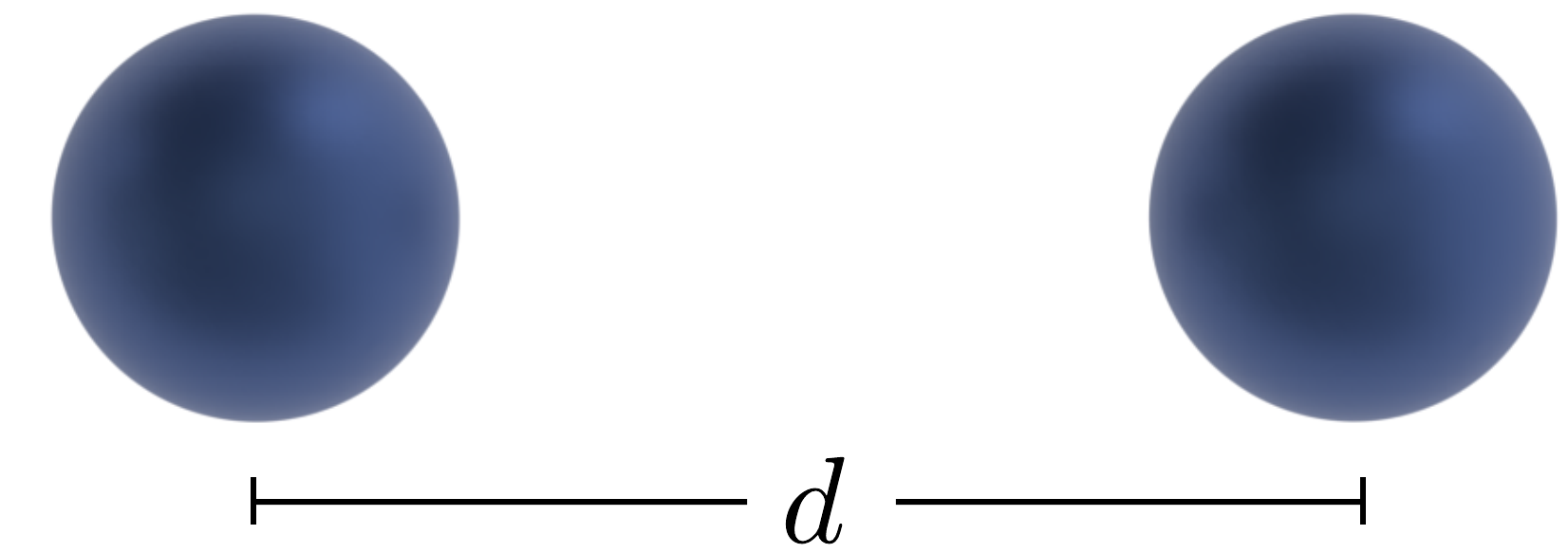}}
        \caption{} 
    \end{subfigure}
    \hfill
    \begin{subfigure}[b]{0.32\textwidth}   
        \centering 
        \includegraphics[width=\textwidth]{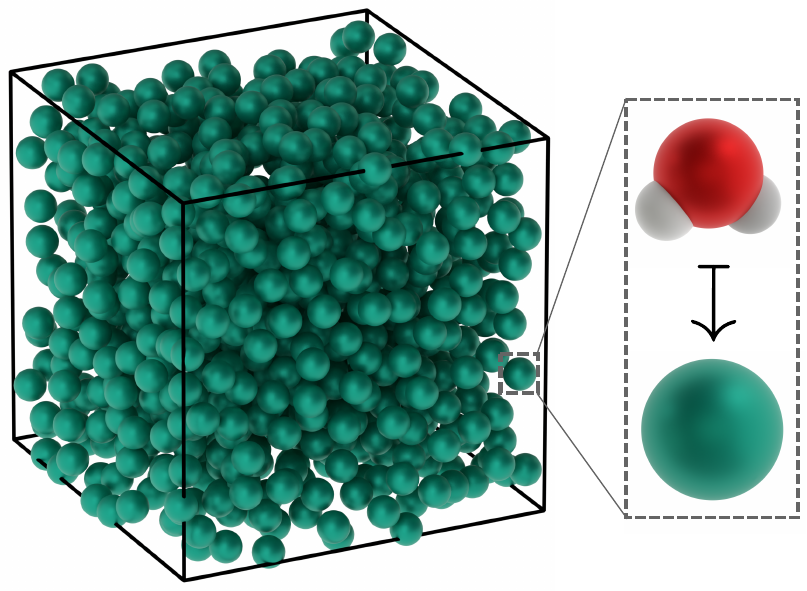}
        \caption{}
    \end{subfigure}
    \hfill
    \begin{subfigure}[b]{0.32\textwidth}   
        \centering 
        \includegraphics[width=\textwidth]{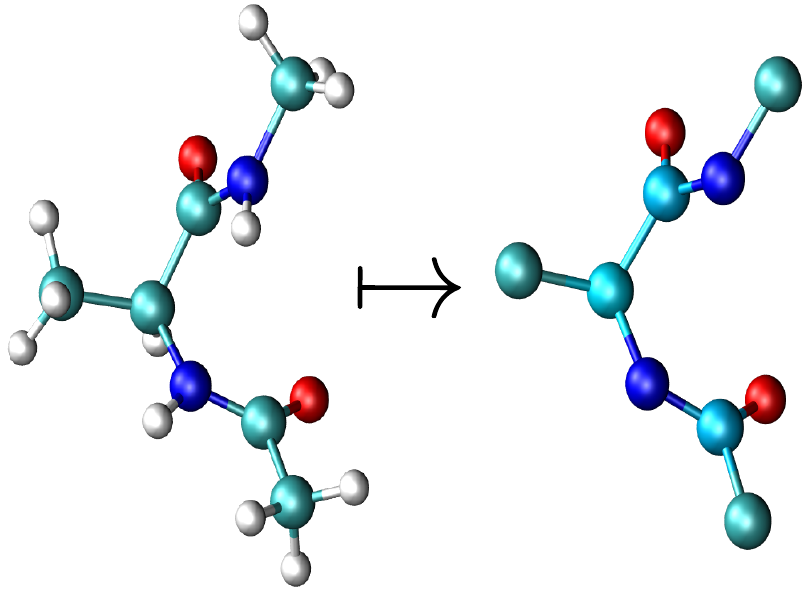}
        \caption{}
    \end{subfigure}
    \caption{Visualization of numerical test case systems: ($a$) Lennard Jones potential, ($b$) coarse-grained liquid water (adapted from Ref.~\citenum{Thaler_2021}), ($c$) coarse-grained alanine dipeptide.}
    \label{fig:system_viz}
\end{figure}

We present three examples (fig. \ref{fig:system_viz}) to distinguish between different sources of uncertainty:
A LJ toy example features epistemic uncertainty only, while the following two CG systems include a significant amount of aleatoric uncertainty. We additionally show the effects of systematic uncertainty for liquid water and for alanine dipeptide.

\subsection{Lennard Jones Potential}

We learn a LJ potential ($\sigma_\mathrm{LJ}, \epsilon$) with a
pairwise additive NN potential to benchmark the scalable UQ methods against a HMC scheme.
\begin{figure}
    \centering
    \includegraphics[width=\textwidth]{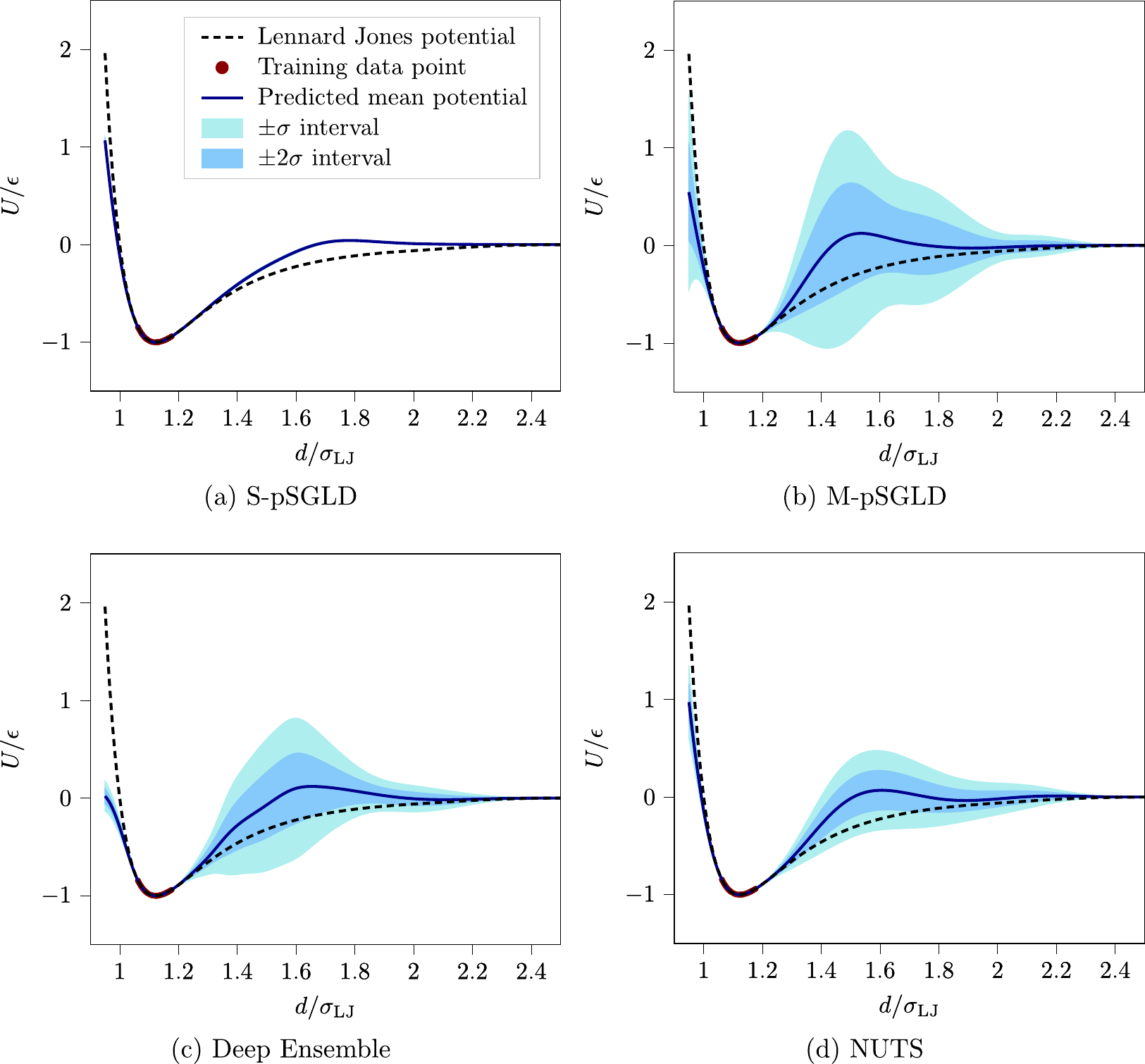}
    \caption{Distribution of NN potentials. Predicted mean potential with $\pm \sigma$ and $\pm 2\sigma$ intervals of the single chain pSGLD ($a$), the multi-chain pSGLD ($b$), the Deep Ensemble method ($c$) and the multi-chain No-U-Turn Sampler (NUTS, $d$), compared to the Lennard Jones reference.}
    \label{fig:toy_data_results}
\end{figure}
As the reference method, we select the No-U-Turn Sampler (NUTS) \cite{Hoffman2014}, which selects the number of HMC integration steps on-the-fly. Additionally, a window adaption warm-up scheme \cite{Gelman2015, Blackjax2020} automatically selects an appropriate mass matrix and step size, such that no hyperparameter tuning is required for the NUTS. 
The NN potential predicts the pairwise potential energy $U(d)$ given pairwise particle distance $d$ and consists of a single hidden layer with 64 neurons and swish activation, where $d$ is represented by six radial Bessel functions \cite{Klicpera2020} with a cut-off $r_\mathrm{cut} =  2.5 \sigma_{LJ}$. We choose a Gaussian prior for weights and biases $p(\mathbf{w}) \sim \mathcal{N}(\mathbf{0}, \mathbf{I})$ and an exponential prior with scale 1 for the aleatoric uncertainty $p(\sigma_\mathrm{H})$.
For all Bayesian methods, we sample 100 models from the Bayesian posterior ($\mathcal{T}=1$), evenly distributed over all considered Markov chains. 
The training data consists of 100 randomly drawn training data points from the well of the LJ potential $d/\sigma_{\mathrm{LJ}} \in [1.0615, 1.1800]$ (supplementary fig.~1).
Additional technical details are provided in supplementary methods~2.

In the following, we benchmark pSGLD with a single chain (S-pSGLD), pSGLD with 10 chains (M-pSGLD) and a Deep Ensemble consisting of 10 NNs against a 10 chain NUTS. The obtained mean potentials and corresponding standard deviation intervals are visualized in fig. \ref{fig:toy_data_results}.
The mean potentials of all considered methods fit the LJ potential very well where training data were generated: On held-out data within the training interval, we obtained low root-mean-squared errors ($\mathrm{RMSE}/\epsilon$) of $0.011$ (S-pSGLD), $0.014$ (NUTS), $0.023$ (M-pSGLD), and $0.025$ (Deep Ensemble).

Bayesian methods estimate the scale of the aleatoric uncertainty to $\sigma_\mathrm{H} / \epsilon \approx 10^{-3}$.
Such low estimates are expected given that the aleatoric uncertainty of the LJ data set is zero and the NN potential has sufficient capacity to interpolate the training data.
Hence, we neglect the contribution of the aleatoric uncertainty in the following uncertainty predictions and only show the epistemic uncertainty.
The S-pSGLD method samples a single posterior mode only, yielding highly overconfident potential energy predictions outside the training interval. By contrast, the other methods using multiple randomly initialized models sample multiple posterior modes, such that they can capture a significant amount of epistemic uncertainty.
Accordingly, the obtained credible intervals include the reference potential across a broad range of distances.
Both M-pSGLD and the Deep Ensemble method provide good approximations to the NUTS reference distribution. However, compared to the NUTS reference, the Deep Ensemble method underestimates uncertainty at short distances and M-pSGLD overestimates uncertainty at medium distances.

All UQ methods sampling multiple modes exhibit a similar shape of the predicted epistemic uncertainty. Local uncertainty maxima are located between $1.4 \sigma_\mathrm{LJ} < d < 1.8 \sigma_\mathrm{LJ}$ and the uncertainty significantly increases for $d < 0.9 \sigma_\mathrm{LJ}$. This uncertainty shape is the result of the NN potential architecture and the location of the training data set: On the one hand, the radial Bessel representation \cite{Klicpera2020} of $d$ smoothly shrinks the NN potential towards 0 at $r_\mathrm{cut}$. On the other hand, the training data constrains the potential for $1.06 \sigma_\mathrm{LJ} < d < 1.18 \sigma_\mathrm{LJ}$. Hence,
these results are consistent with the expectation that the epistemic uncertainty should increase with the distance from
points that constrain the potential.

\begin{figure}[htbp]
    \centering
    \includegraphics[width=\textwidth]{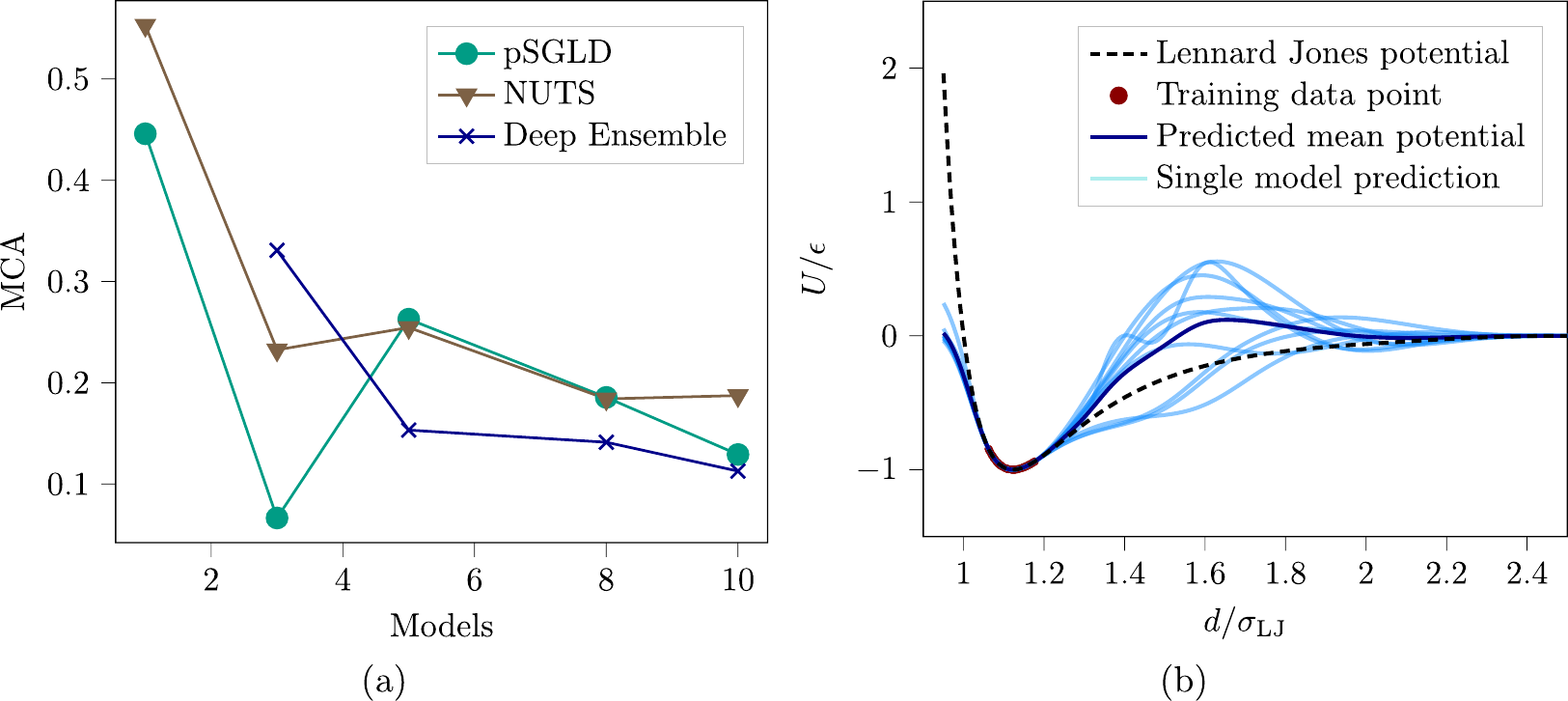}
    \caption{Posterior mode analysis. ($a$) miscalibration area (MCA) of the No-U-Turn Sampler (NUTS), the pSGLD, and the Deep Ensemble methods as a function of the number of randomly initialized models. The MCA includes both within and out-of-distribution test data. ($b$) All predicted potentials of the Deep Ensemble method with resulting mean compared to Lennard Jones reference.}
    \label{fig:LJ_MCA_CHAINS}
\end{figure}

We further investigate the effect of the number of randomly initialized models (number of Markov chains for MCMC) on UQ quality. The miscalibration area (MCA) \cite{Tran2020, Chung2021, Hirschfeld2020} quantifies the agreement between the predicted standard deviation and the true error.
For all methods, the MCA shows a decreasing trend with increasing number of randomly initialized models (fig. \ref{fig:LJ_MCA_CHAINS} $a$).
This reflects the importance of sampling multiple posterior modes for robust UQ \cite{Wilson2020}, which can be achieved comparatively easily by exploiting the strong decorrelation effect of random NN initializations \cite{Fort2019, Wilson2020}.
Different posterior modes represent different potentials, all of which are consistent with the training data, but differ significantly where there is no data available, thus capturing the epistemic uncertainty (fig. \ref{fig:LJ_MCA_CHAINS} $b$).

The inability to sample multiple posterior modes using a single Markov chain is not unique to pSGLD.
A single chain of the NUTS also samples a single posterior mode only and the captured epistemic uncertainty increases with additional chains (see also supplementary fig.~2).
This suggests that sampling multiple posterior modes with a single Markov chain is difficult to achieve when training NN potentials, even for sophisticated posterior exploration schemes.
Finally, we note that by artificially fixing $\sigma_\mathrm{H}$ to a large value as in Ref. \citenum{Kahle2022}, the single chain NUTS predicts large epistemic uncertainty outside the training interval (supplementary fig.~3). However, this comes at the cost of a larger error within the training interval ($\mathrm{RMSE} / \epsilon = 0.044$ for $\sigma_\mathrm{H} / \epsilon=0.05$) given that models with poorer fit also appear probable due to the allegedly large aleatoric noise in the data.

\subsection{Coarse-grained Liquid Water}
We apply pSGLD and the Deep Ensemble method to CG liquid water, a classic benchmark problem, to test their respective performance both within the training distribution as well as under distribution shift. The reference data consists of 100 cubic boxes of length $l=3.129$ nm containing 1000 water molecules each, sampled every 1 ps from the TIP4P/2005 \cite{Abascal2005} model at a temperature $T_\mathrm{ref}=298$~K, resulting in a pressure $p_\mathrm{ref} = -6.2 \ \mathrm{MPa}$. We divide the data into training, validation and test with a 80\%-8\%-12\% split. Each water molecule is modeled by a CG particle positioned at its center of mass.

We select the repulsive part of the LJ potential as prior potential
\begin{equation}
\label{eq:prior_water}
    U^\mathrm{prior}(\mathbf{R}) = \sum_{i=1}^{N_\mathrm{pair}} \epsilon_\mathrm{w} \left( \frac{\sigma_\mathrm{w}}{d_i} \right)^{12} \ ,
\end{equation}
with $\epsilon_\mathrm{w} = 1\ \mathrm{kJ}/\mathrm{mol}$ and $\sigma_\mathrm{w} = 0.3165$~nm, where $\sigma_\mathrm{w}$ corresponds to the length scale of the SPC water model \cite{Berendsen1981}. This corresponds to the $U^\mathrm{prior}$ used in our previous works, where we found DimeNet++ results to be insensitive to the specific prior potential chosen \cite{Thaler_2021, Thaler2022}. We account for the thermodynamic state point dependency of the PMF \cite{Noid2008, Chaimovich2009, Potestio2014} by augmenting the edge embedding of DimeNet++ by two learnable 16 dimensional vector, one multiplied and one divided by $k_\mathrm{B}T$. This dual embedding ensures that the temperature dependency effect vanishes for neither high nor low $k_\mathrm{B}T$.

The models are trained with a batch size of 5 boxes, an initial learning rate $a=5\cdot 10^{-4}$ and a polynomial learning rate decay schedule (eq.~\eqref{eq:lr_schedule}) with $\gamma=0.55$. We generate a Deep Ensemble of 8 models and train each for 100 epochs with the Adam \cite{Kingma2015} optimizer with default parameters. For each training trajectory, we select the model parameters with the smallest validation loss, giving the Deep Ensemble method a slight advantage in terms of data usage over the Bayesian methods.
For Bayesian modeling, we select a prior distribution $p(\sigma_\mathrm{H}) \sim \Gamma(5, 27)$, incorporating the prior knowledge that $\sigma_\mathrm{H} > 0$ due to the noise from the non-injective CG mapping \cite{Wang2009}. By default, we select a posterior temperature $\mathcal{T}=0.01$. Each pSGLD chain is run for 10000 epochs, 8000 of which are discarded as burn-in. We randomly subsample the remaining models such that a total of 40 models are selected, evenly distributed over all available chains (8 chains for M-pSGLD, 1 chain for S-pSGLD).
One chain of the M-pSGLD method yielded poor potentials and we omitted it for a more balanced comparison.

First, we evaluate the mean force predictions on the test data. The Deep Ensemble method with a RMSE of $135.8\ \mathrm{kJ / (mol\ nm)}$ is more accurate than S-pSGLD and M-pSGLD with $\mathrm{RMSE} = 137.2\ \mathrm{kJ / (mol\ nm)}$ and $\mathrm{RMSE} = 136.6\ \mathrm{kJ / (mol\ nm)}$, respectively.
We were unable to find a set of pSGLD hyperparameters that closed the error differential to the Deep Ensemble method. 
The force error is dominated by the large aleatoric uncertainty, which is estimated as $\sigma_\mathrm{H} = 136.6\ \mathrm{kJ / (mol\ nm)}$ by S-pSGLD.

\begin{figure}
    \centering
    \includegraphics[width=0.5\textwidth]{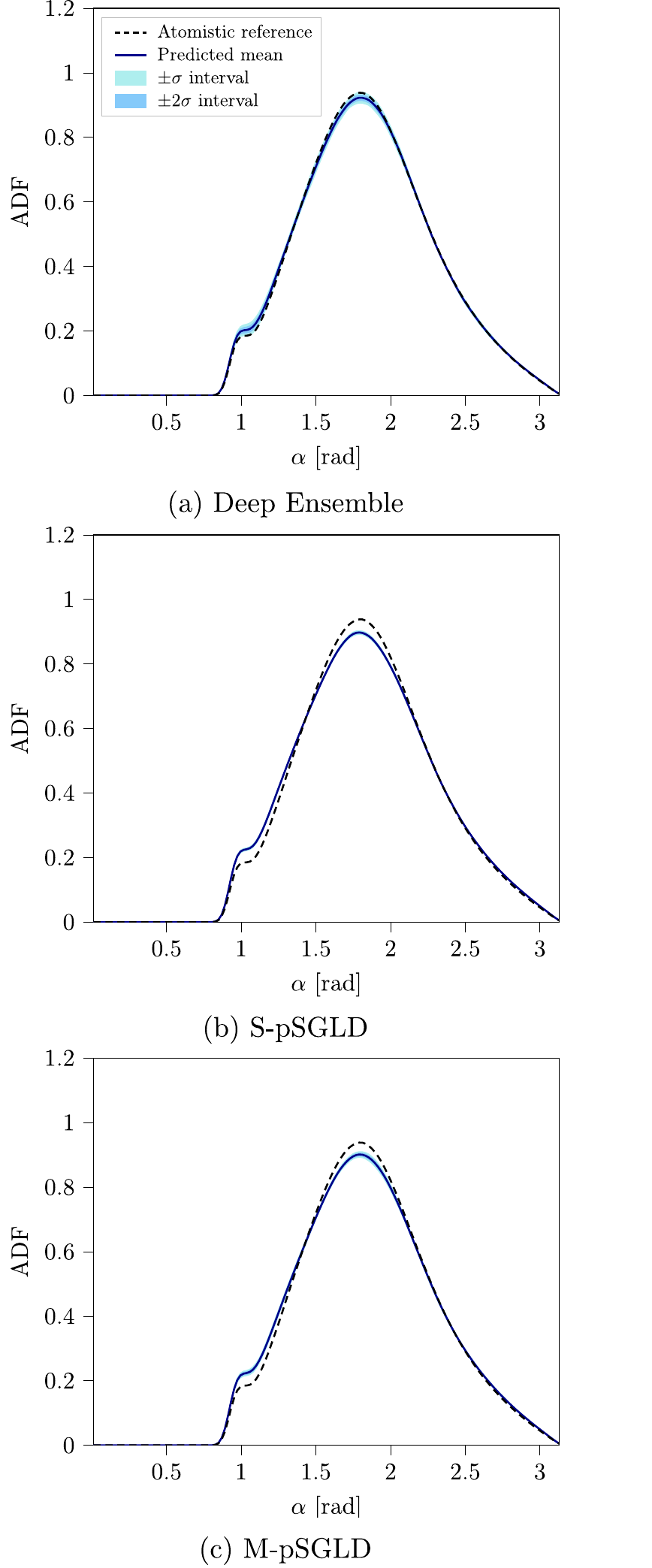}
    \caption{Angular distribution functions (ADF) at $T = T_\mathrm{ref}$. Resulting mean ADF with $\pm \sigma$ and $\pm 2\sigma$ intervals as predicted by the Deep Ensemble method ($a$), the single chain pSGLD ($b$) and the multi-chain pSGLD ($c$) schemes at a temperature $T = T_\mathrm{ref}$, compared to the atomistic reference.}
    \label{fig:water_within_dist}
\end{figure}

We run CG MD simulations at a temperature $T=T_\mathrm{ref}$ to investigate the resulting observables without a distribution shift.
All CG MD simulations use a time step of 2 fs and are equilibrated for 10 ps, followed by 100 ps of production, where a state is retained every 0.1 ps.
The CG MD simulation averages over the aleatoric uncertainty resulting from the  CG mapping. Consequently, the predicted standard deviation of observables $\sigma$ includes the epistemic uncertainty as well as a small amount of MD sampling uncertainty due to finite trajectory lengths.
Fig. \ref{fig:water_within_dist} shows the resulting distributions of
angular distribution functions (ADF). The mean prediction of the Deep Ensemble method matches the AT reference well and is slightly more accurate than the S-pSGLD and M-pSGLD schemes, reflecting the lower test set RMSE. Additionally, the $2\sigma$ credible interval of the Deep Ensemble method covers the AT reference, and areas with higher uncertainty correspond to areas with larger error. M-pSGLD captures slightly more variance than S-pSGLD, but both schemes are overconfident.
The overconfidence of M-pSGLD seems to be primarily attributable to a larger deviation of the predicted mean ADF from the reference curve (in line with the larger test set RMSE) and only secondarily to less captured epistemic uncertainty compared to the Deep Ensemble method.
The conclusions drawn from the RDF predictions are identical (supplementary fig.~4).

We investigate the impact of the Gaussian prior for weights and biases by retraining the NN potential with an improper uniform distribution. The obtained results for S-pSGLD are largely identical to the Gaussian prior case (supplementary fig.~5).
However, the Gaussian prior appears to improve the learning robustness: With the uniform distribution, a total of 4 M-pSGLD Markov chains yielded models with large errors in the mean predictions, compared to only a single Markov chain with the Gaussian prior.
Having verified that neither of the priors $p(\mathbf{w})$ and $p(\sigma_\mathrm{H})$ are
too restrictive, we hypothesize that the higher RMSEs of the pSGLD schemes may be the result of the training, where the coupling of learning rate and additive random noise might impede convergence to models with the highest likelihood.

\begin{figure}
    \centering
    \includegraphics[width=0.6\textwidth]{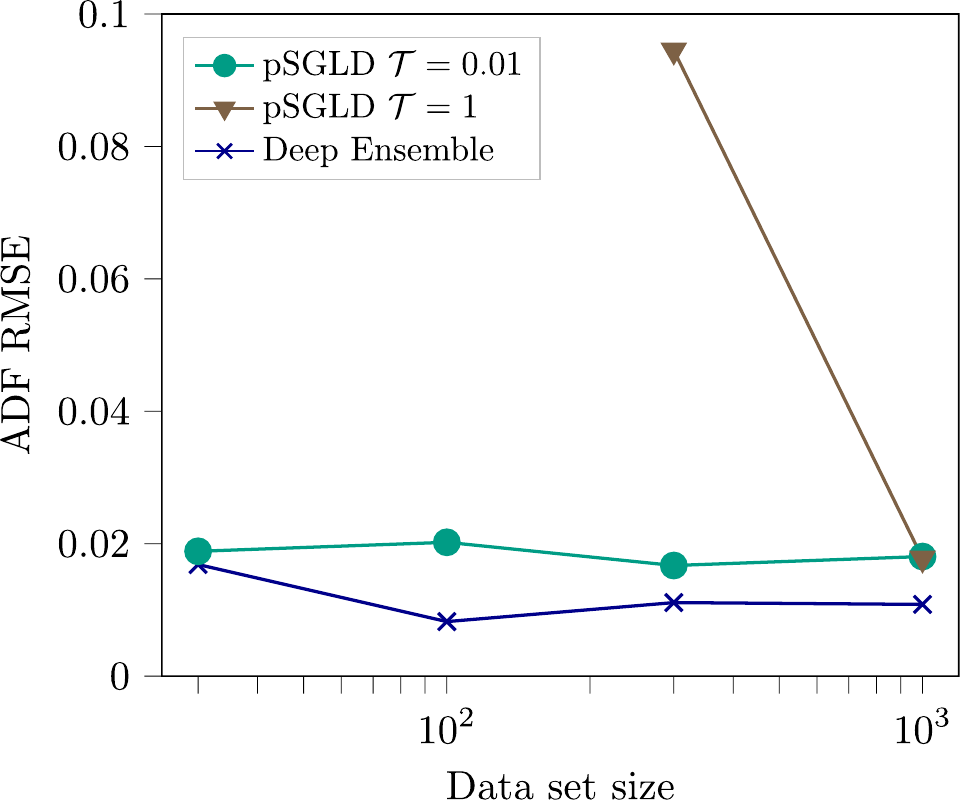}
    \caption{Cold posterior effect. Root mean squared error (RMSE) of the mean predicted angular
    distribution function (ADF) at $T=T_\mathrm{ref}$ of the Deep Ensemble method and the single chain pSGLD scheme with $\mathcal{T}=1$ and $\mathcal{T}=0.01$ for different data sizes. Note that pSGLD $\mathcal{T}=1$ yields unstable MD simulations for data sizes of 30 and 100 boxes.}
    \label{fig:cold_posterior}
\end{figure}

Next, we investigate the impact of the posterior temperature $\mathcal{T}$ on pSGLD models (fig. \ref{fig:cold_posterior}). 
With the Bayesian posterior $\mathcal{T} = 1$, S-pSGLD requires a large data set set (1000 boxes) to sample accurate models. For a medium data set size (300 boxes), the obtained models are highly inaccurate compared to using the cold posterior $\mathcal{T}=0.01$. For smaller data set sizes, models sampled with the Bayesian posterior result in unstable CG MD simulations. By contrast, the cold posterior allows to sample accurate models with only a fraction of the data (30 boxes).
Moreover, the accuracy of pSGLD models hinges on a sufficient amount of burn-in epochs to reduce the learning rate (supplementary fig.~6).
Consequently, the pSGLD schemes require significantly more computational training effort in this example than the Deep Ensemble method. Still, the Deep Ensemble method yields more accurate models for all data set sizes considered in fig. \ref{fig:cold_posterior}.

\begin{figure}
    \centering
    \includegraphics[width=0.465\textwidth]{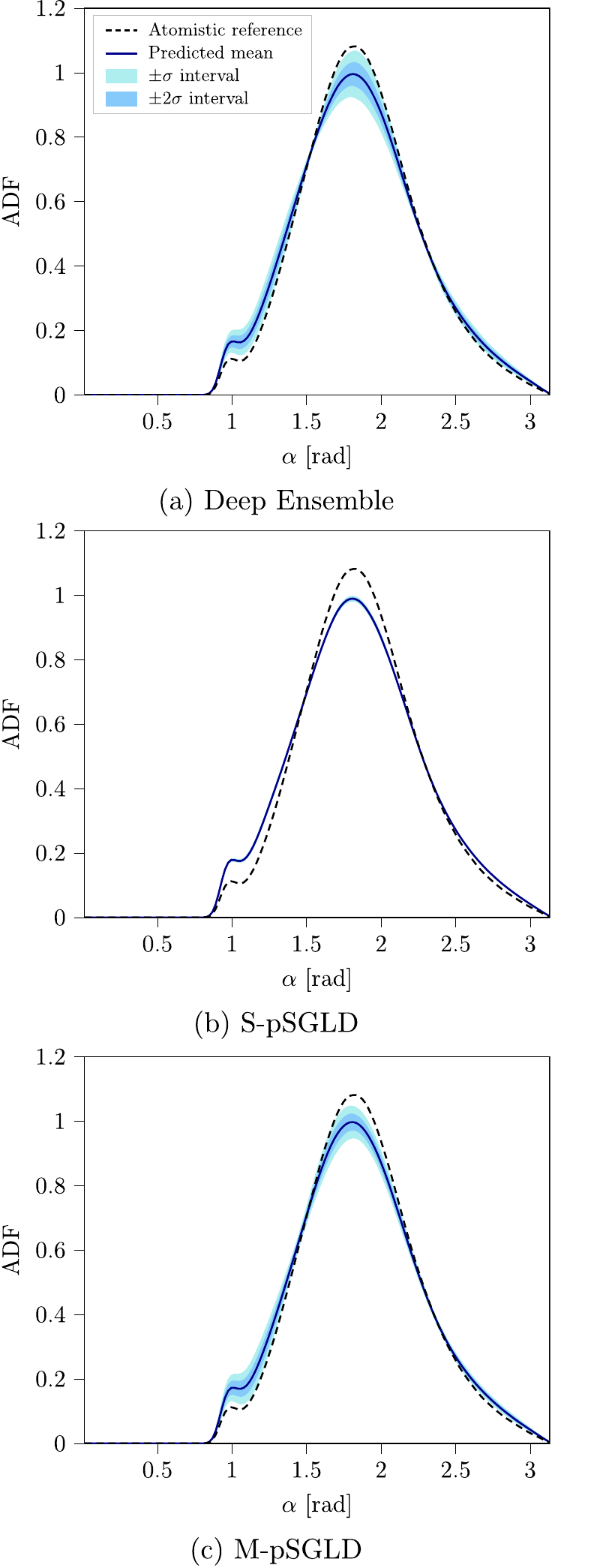}
    \caption{Out-of-distribution angular distribution functions (ADF) at $T = 260$ K. Resulting mean ADF with $\pm \sigma$ and $\pm 2\sigma$ intervals as predicted by the Deep Ensemble ($a$), the single chain pSGLD ($b$) and the multi-chain pSGLD ($c$) schemes at a temperature $T = 260$ K, compared to the atomistic reference.}
    \label{fig:water_out_of_dist}
\end{figure}

To test the quality of UQ under distribution shift, we apply the obtained models at a temperature $T = 260$ K. The mean predictions of the considered UQ schemes are very similar to each other and, as expected, deviate from the respective TIP4P results (fig. \ref{fig:water_out_of_dist}).
While S-pSGLD results in highly overconfident predictions, both M-pSGLD and the Deep Ensemble method provide accurate credible intervals, with a slight advantage for the latter.
The predicted RDFs allow for identical conclusions (supplementary fig.~7).
The accurate ADF credible intervals we obtained with the M-pSGLD and the Deep Ensemble method stand in contrast to previous findings with a 2-body cubic spline model \cite{Thaler2022b}: Given that the 2-body spline cannot model many-body effects, uncertainty with respect to 3-body interactions cannot be captured in the epistemic uncertainty. This model misspecification results in systematic uncertainty not included in the credible interval.
This highlights the advantage of the many-body capabilities of NN potentials in a UQ context.

\begin{figure}
    \centering
    \includegraphics[width=\textwidth]{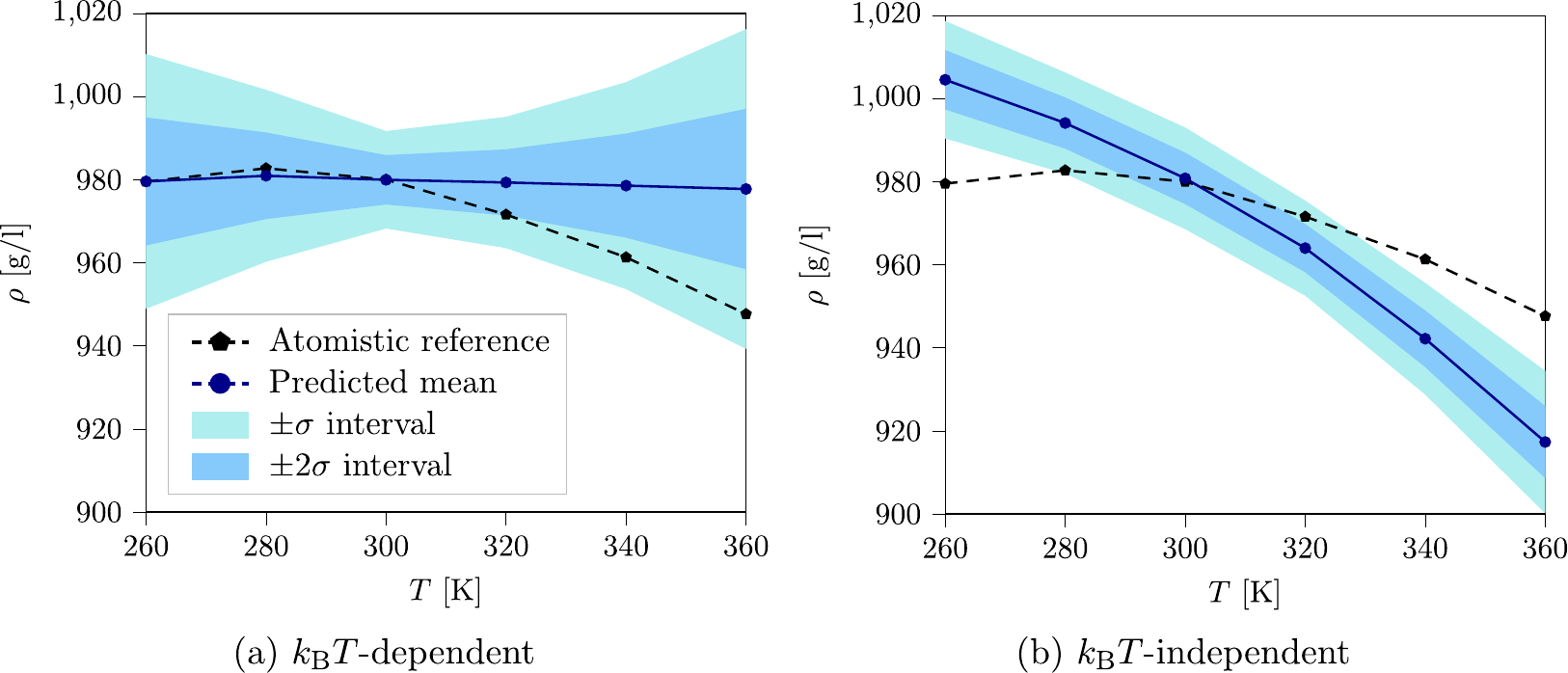}
    \caption{Water density profile. Resulting mean density $\rho$ at pressure $p=1$ bar with $\pm \sigma$ and $\pm 2\sigma$ intervals as predicted by the Deep Ensemble method using the $k_\mathrm{B}T$-dependent reference model ($a$) and the same model without the $k_\mathrm{B}T$-dependent edge embedding ($b$), compared to the atomistic reference.}
    \label{fig:density}
\end{figure}

Finally, we study the impact of the $k_\mathrm{B}T$-dependent edge embedding. In the first step, we match the AT reference pressure at $T_\mathrm{ref}$ during the FM training (details in supplementary methods~3).
Using the Deep Ensemble method, we then compute the density $\rho$ as a function of temperature with and without the $k_\mathrm{B}T$-dependent embeddings (fig. \ref{fig:density}). 
As desired, the credible interval includes the AT reference and the uncertainty increases with the distance from the training temperature $T_\mathrm{ref}$ for the $k_\mathrm{B}T$-dependent model.
By contrast, without $k_\mathrm{B}T$-dependent embedding, the predicted uncertainty barely increases with the distance from $T_\mathrm{ref}$, resulting in overconfident predictions due to model misspecification.
Given that the $k_\mathrm{B}T$ dependence enables a broader range of outcomes at $T \neq T_\mathrm{ref}$, the mean predictions also change significantly and yield smaller errors further away from $T_\mathrm{ref}$.
These results highlight the potency of scalable UQ methods to quantify errors resulting from applying CG models at different thermodynamic state points than during training.

\subsection{Coarse-grained Alanine Dipeptide}
We consider the benchmark problem of learning the free energy surface (FES) of alanine dipeptide \cite{Pettitt1985, Tobias1992}, which has recently been shown to be a challenging task for NN potentials trained via FM \cite{Fu2022, Thaler2022}. Here, we investigate the sources of these challenges using the scalable UQ toolbox. We build on the computational setup of our previous study \cite{Thaler2022}: The CG map retains all 10 heavy atoms of alanine dipeptide, dropping hydrogen atoms and water molecules. The CG particles modeling $\mathrm{CH}_3$, CH and C are encoded as different particle types. The training data set consists of a 100~ns AT trajectory at $T_\mathrm{ref} = 300$~K, which is subsampled to $5 \cdot 10^5$ data points by retaining a state every $0.2$~ps. The first 80~ns form the training set, the subsequent 8~ns the validation set. To counteract the instability of DimeNet++ in CG MD simulations of alanine dipeptide \cite{Fu2022}, we add a prior potential $U^\mathrm{prior}(\mathbf{R})$ (eq.~\eqref{eq:prior_addition}) that consists of harmonic bonds and angles, as well as proper dihedrals. 
For more technical details on $U^\mathrm{prior}$ and the AT reference data, we refer to our previous work \cite{Thaler2022}.

We train all models with an initial learning rate $a= 10^{-3}$ and a polynomial learning rate decay schedule (eq.~\eqref{eq:lr_schedule}) with $\gamma=0.55$, as well as a batch size of 512 configurations.
The 8 models of the Deep Ensemble method are trained for 1000 epochs, using the Adam \cite{Kingma2015} optimizer with default parameters. For each training trajectory, we select the model parameters with the smallest validation loss.
pSGLD chains are run for 3000 epochs, with the first 2500 epochs discarded as burn-in.
We randomly subsample the remaining models such that a total of 40 models are selected, evenly distributed over all available chains (8 chains for M-pSGLD, 1 chain for S-pSGLD). We select a prior distribution $p(\sigma_\mathrm{H}) \sim \Gamma(10, 40)$ and a posterior temperature $\mathcal{T}=0.05$.

We initially evaluate performance of the considered methods on the test set: 
The S-pSGLD ($\mathrm{RMSE} = 414.12\ \mathrm{kJ / (mol\ nm)}$), M-pSGLD ($\mathrm{RMSE} = 414.01\ \mathrm{kJ / (mol\ nm)}$) and Deep Ensemble methods ($\mathrm{RMSE} = 413.84\ \mathrm{kJ / (mol\ nm)}$) yield very similar accuracy, with a slight advantage for the Deep Ensemble method.
This is in line with the aleatoric uncertainty scale $\sigma_\mathrm{H} = 414.69$, estimated by S-pSGLD.

We perform a 100 ns CG MD production simulation for all sampled models in order to compute the FES. To obtain the same number of trajectories as with the pSGLD schemes, each of the 8 Deep Ensemble models generates 5 trajectories, all starting from different initial states.
Despite using a prior potential, some models became stuck in unphysical potential energy "holes" \cite{Van2022}, i.e. deep potential energy minima in rarely sampled phase-space regions,
which also led to instability in some cases.
These potential energy holes might be avoided by employing better prior potentials or by incorporating MD simulations into training, e.g. via active learning \cite{Smith2018, Van2022} or alternative training schemes such as relative entropy (RE) minimization \cite{Shell2008, Chaimovich2011, Thaler2022}.
We note that using the Bayesian posterior $\mathcal{T} = 1$ significantly increased the number of unphysical trajectories (tested for S-pSGLD). For $\mathcal{T} = 1$, we observed results of comparable quality to $\mathcal{T} = 0.05$ only when increasing the data set to $1 \mu s$.

\begin{figure}
    \centering
    \includegraphics[width=\textwidth]{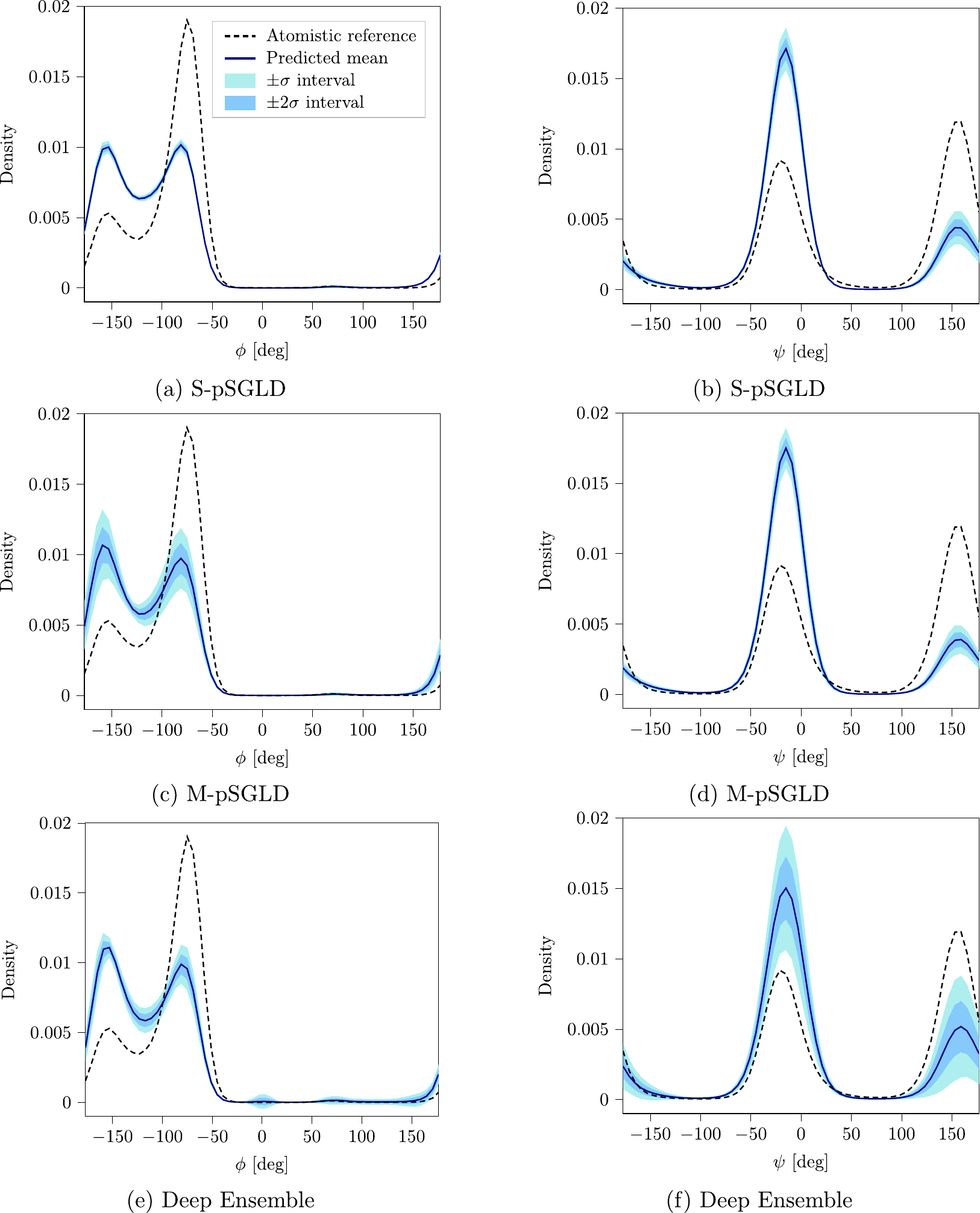}
    \caption{Dihedral angle density histograms. Resulting mean distribution of dihedral angles $\phi$ (left column) and $\psi$ (right column) with $\pm \sigma$ and $\pm 2\sigma$ intervals as predicted by the single chain pSGLD ($a$, $b$), the multi-chain pSGLD ($c$, $d$) and the Deep Ensemble ($e$, $f$) methods, compared to the atomistic reference.}
    \label{fig:alanine_density}
\end{figure}

First, we investigate UQ results after removing unphysical trajectories that mainly sampled configurations in a potential energy hole. To this end, we removed 1, 7 and 13 trajectories from the S-pSGLD, the M-pSGLD and the Deep Ensemble methods, respectively.
The resulting means and standard deviations of the dihedral angles $\phi$ and $\psi$ \cite{Pettitt1985, Tobias1992} are shown in fig. \ref{fig:alanine_density}. The mean predictions of S-pSGLD and M-pSGLD are very similar, but -- consistent with the examples above -- S-pSGLD significantly underestimates the epistemic uncertainty. The Deep Ensemble method yields similar mean predictions in $\phi$ and a slightly improved mean prediction in $\psi$. The M-pSGLD predicts larger epistemic uncertainty in $\phi$, while the Deep Ensemble method predicts larger uncertainty in $\psi$.
However, all considered methods show that the epistemic uncertainty is not sufficiently large to fully account for the deviation from the AT reference in this case.

To contextualize this result, we replace the FM training by RE minimization. Given that the RE model trained on the same data set can match the AT FES accurately \cite{Thaler2022}, insufficient model capacity is not the main limiting factor.
Additionally, a poor approximation of the posterior by the considered UQ methods, which would result in an incorrect size of the predicted credible interval, can also be ruled out:
The posterior probability ratio of the RE model and the last model sampled by the S-pSGLD FM scheme is $p(\theta_\mathrm{RE}|\mathcal{D}) / p(\theta_\mathrm{FM}|\mathcal{D}) = \mathrm{e}^{-25872}$. This is in line with previous findings showing that the error on held-out force data is smaller for FM than for RE for alanine dipeptide \cite{Thaler2022}. Given that the posterior probability ratio of the RE model is numerically zero, UQ schemes with a FM-based posterior cannot sample the RE model.

Multiple mechanisms may contribute to the comparatively weak FES prediction of FM models.
First, the FES of a FM model is sensitive to predictions in sparsely resolved transition regions \cite{Thaler2022}. Second, if a CG MD simulation is able to reach unphysical phase-space regions,
sampling such configurations yields an erroneous FES.
Both of these mechanisms result in very large epistemic uncertainty.
We empirically show this effect when we include trajectories that sampled potential energy holes in the evaluation of the FES distribution (supplementary fig.~8).
In particular, the predicted credible intervals of M-pSGLD and the Deep Ensemble method mostly cover the reference FES in this case. Hence, these UQ methods can signal to practitioners that the obtained results are not yet trustworthy.

Rather, more data needs to be generated to further constrain the learned models.
We increased the data set size by factor 10 by generating an AT trajectory of 1 $\mu$s length.
Interestingly, simply generating more Boltzmann-distributed training data did not solve the potential energy hole problem, nor did it significantly reduce the deviation of the mean prediction when neglecting trajectories stuck in potential energy holes (supplementary fig.~9).
Hence, generating more diverse, non-Boltzmann distributed data sets, e.g. via enhanced sampling schemes \cite{Schneider1989, Barducci2011} or active learning \cite{Smith2018, Van2022}, seems to be a more promising approach.

The remaining deviation of the mean prediction from the reference FES that is not captured by the predicted epistemic uncertainty (fig. \ref{fig:alanine_density}) suggests that other, likely systematic sources of error exist.
For finite model capacity, a systematic difference between FM and RE minimization is the objective function:
RE training minimizes the difference between the potential energy surfaces of the AT and CG models \cite{Chaimovich2010}, which is directly related to the FES.
In contrast, the optimum of a force-based training objective might trade off accuracy in the FES for improved accuracy in other (e.g. thermodynamic \cite{Thaler2022}) observables, resulting in systematic uncertainty in the predicted FES.
Additionally, numerical errors introduced by the CG MD simulation, similar to the shadow Hamiltonian effect \cite{Toxvaerd1994, Toxvaerd2013}, can be corrected for by RE minimization \cite{Thaler2022}.
In FM models, these numerical errors manifest as unquantifiable systematic uncertainty.
However, for a comprehensive analysis of the relative impact of each error mechanism, further research is needed.

\section{Discussion and Conclusion}
Our results show that M-pSGLD is well suited to estimate the epistemic uncertainty of MD observables. This method enables fully-Bayesian UQ for NN potentials. All experiments highlight the importance of sampling multiple posterior modes. Exploiting the strong decorrelation effect of multiple random NN initializations via multiple Markov chains is an effective means to this end.
In the graph NN examples, cold posteriors proved beneficial to sample both stable and accurate models, reducing the required amount of training data significantly.
Hence, we found the number of Markov chains to be the most important additional M-pSGLD hyperparameter, followed by the posterior temperature, the prior distributions and the number of samples per chain.

Both Deep Ensemble and the M-pSGLD methods provided good approximations to the epistemic uncertainty estimated by the NUTS \cite{Hoffman2014} in the LJ example.
In addition, the Deep Ensemble method yielded similar UQ quality to M-pSGLD, although it required less training and hyperparameter tuning effort.
We found no evidence that the Deep Ensemble method was prone to overconfident predictions, contrasting prior research in an active learning setting \cite{Kahle2022}. Instead, our results suggest that the Deep Ensemble method quantifies epistemic uncertainty effectively, both within and out of the training distribution.

M-pSGLD promises accurate UQ by leveraging the complementary benefits from sampling multiple posterior modes and additional Bayesian exploration of each mode \cite{Fort2019, Wilson2020, Izmailov2021}.
However, further research into SG-MCMC schemes is required before routine application in practice:
In our experiments, a single MCMC chain (both pSGLD \cite{Welling2011} and NUTS \cite{Hoffman2014}) sampled a single posterior mode only.
Hence, the development of methods that sample multiple posterior modes with a single chain, e.g., by leveraging cyclical step size schedules \cite{Zhang2020} or parallel tempering \cite{Deng2020}, is important.
Additionally, automatic hyperparameter tuning with a computationally efficient metric could improve the SG-MCMC efficiency; e.g., the popular Stein's discrepancy \cite{Gorham2015} scales quadratically with the data set size \cite{Nemeth2021}.
Finally, recent SG-MCMC samplers such as AMAGOLD \cite{Zhang2020} or SGGMC \cite{Garriga2020} include infrequent Metropolis-Hastings acceptance steps to avoid the bias of SGLD \cite{Welling2011, Li2016}. Consequently, these samplers use constant learning rates, which may counteract the increased training time of SGLD that results from its small learning rate requirement \cite{Teh2016, Zhang2020}.

We observed a clear cold posterior effect \cite{Wenzel2020} in our experiments with the graph NN potential.
For image classification tasks, cold posteriors have demonstrated superior performance in practice \cite{Zhang2019cyc,Heek2020,Wenzel2020}. However, this performance increase is mainly attributed to data augmentation \cite{Izmailov2021}, which increases the effective data size without increasing the data size considered in the likelihood. Analogously, the effective data size might be underestimated by the likelihood in eq.~\eqref{eq:Likelihood_MD}, which the cold posterior may correct for: When learning the potential energy of a molecular state, the effective data size is clearly larger than a single data point. For instance, in the case of a pairwise additive potential, the effective data size corresponds to the total number of particle pairs within a cutoff. For FM, the data size per box considered in the likelihood in eq.~\eqref{eq:Likelihood_MD} equals 3 times the number of CG particles, but whether the effective data size exceeds this value is less clear. More research into the nature of the cold posterior effect is required -- ideally resulting in likelihood formulations that better consider the effective data size.

Our results corroborate that for successful UQ, a sufficiently large hypothesis space is necessary:
Effects describable by the model can be quantified reliably as epistemic uncertainty, but effects beyond the model capacity become hard to quantify systematic uncertainties \cite{Hullermeier2021, Gal2022}.
For instance, if a potential lacks important many-body interactions or a CG model lacks state-point dependency, the resulting uncertainty estimates are overconfident.
Consequently, NN potentials are attractive models in a UQ context, given that they model many-body interactions inherently.

To obtain uncertainty estimates for MD observables, we performed a dedicated MD simulation for every sampled NN potential.
This approach is rigorous, as both epistemic uncertainty and MD sampling uncertainty are captured \cite{Wan2021}, but also computationally expensive. The computational effort for MD simulations scales linearly with the number of sampled potentials, but the simulations can be parallelized.
Distilling the mean potential energy prediction into a single model via student-teacher \cite{Korattikara2015, Li2016, Wang2021} learning could improve computational efficiency. With this approach, one could obtain uncertainty estimates for time-averaged observables using a single MD simulation and a reweighting scheme \cite{Imbalzano2021}. Concerning the development of computationally more efficient UQ schemes for NN potentials \cite{Zhu2022}, we have demonstrated that both M-pSGLD and the Deep Ensemble method can serve as reliable baseline schemes. 
Efficient UQ schemes may pave the way for more reliable MD simulations based on NN potentials to support simulation-based decision-making in health care and material science industries  \cite{Wan2021}.

\begin{acknowledgement}
The authors thank Paul Fuchs for addressing jax-sgmc feature requests.
S.T. acknowledges financial support from the Munich Data Science Institute Seed Fund.
\end{acknowledgement}

\begin{suppinfo}
Background information on the prior potential, training details of the Lennard Jones example and training data visualization, pressure correction scheme, NUTS predictions for different number of Markov chains and fixed $\sigma_\mathrm{H}$, predicted CG water RDFs, ADFs for S-pSGLD with uniform prior over weights and biases, ADF RMSE for different pSGLD Markov chain lengths, dihedral angle density histograms without removing potential energy holes as well as for the 1 $\mu$s data set.
This information is available free of charge via the Internet at \href{https://pubs.acs.org/}{https://pubs.acs.org/}.
\end{suppinfo}

\bibliography{library_RE}

\providecommand{\latin}[1]{#1}
\makeatletter
\providecommand{\doi}
  {\begingroup\let\do\@makeother\dospecials
  \catcode`\{=1 \catcode`\}=2 \doi@aux}
\providecommand{\doi@aux}[1]{\endgroup\texttt{#1}}
\makeatother
\providecommand*\mcitethebibliography{\thebibliography}
\csname @ifundefined\endcsname{endmcitethebibliography}
  {\let\endmcitethebibliography\endthebibliography}{}
\begin{mcitethebibliography}{90}
\providecommand*\natexlab[1]{#1}
\providecommand*\mciteSetBstSublistMode[1]{}
\providecommand*\mciteSetBstMaxWidthForm[2]{}
\providecommand*\mciteBstWouldAddEndPuncttrue
  {\def\EndOfBibitem{\unskip.}}
\providecommand*\mciteBstWouldAddEndPunctfalse
  {\let\EndOfBibitem\relax}
\providecommand*\mciteSetBstMidEndSepPunct[3]{}
\providecommand*\mciteSetBstSublistLabelBeginEnd[3]{}
\providecommand*\EndOfBibitem{}
\mciteSetBstSublistMode{f}
\mciteSetBstMaxWidthForm{subitem}{(\alph{mcitesubitemcount})}
\mciteSetBstSublistLabelBeginEnd
  {\mcitemaxwidthsubitemform\space}
  {\relax}
  {\relax}

\bibitem[Behler and Parrinello(2007)Behler, and Parrinello]{Behler2007}
Behler,~J.; Parrinello,~M. {Generalized neural-network representation of
  high-dimensional potential-energy surfaces}. \emph{Phys. Rev. Lett.}
  \textbf{2007}, \emph{98}, 146401\relax
\mciteBstWouldAddEndPuncttrue
\mciteSetBstMidEndSepPunct{\mcitedefaultmidpunct}
{\mcitedefaultendpunct}{\mcitedefaultseppunct}\relax
\EndOfBibitem
\bibitem[Sch{\"{u}}tt \latin{et~al.}(Dec. 4--9, 2017)Sch{\"{u}}tt, Kindermans,
  Sauceda, Chmiela, Tkatchenko, and M{\"{u}}ller]{Schutt2017}
Sch{\"{u}}tt,~K.~T.; Kindermans,~P.~J.; Sauceda,~H.~E.; Chmiela,~S.;
  Tkatchenko,~A.; M{\"{u}}ller,~K.~R. {SchNet: A continuous-filter
  convolutional neural network for modeling quantum interactions}. Advances in
  Neural Information Processing Systems. Long Beach, CA, USA, Dec. 4--9,
  2017\relax
\mciteBstWouldAddEndPuncttrue
\mciteSetBstMidEndSepPunct{\mcitedefaultmidpunct}
{\mcitedefaultendpunct}{\mcitedefaultseppunct}\relax
\EndOfBibitem
\bibitem[Gilmer \latin{et~al.}(Aug. 6--11, 2017)Gilmer, Schoenholz, Riley,
  Vinyals, and Dahl]{Gilmer2017}
Gilmer,~J.; Schoenholz,~S.~S.; Riley,~P.~F.; Vinyals,~O.; Dahl,~G.~E. {Neural
  message passing for quantum chemistry}. Proceedings of the 34th International
  Conference on Machine Learning. Sydney, Australia, Aug. 6--11, 2017; pp
  1263--1272\relax
\mciteBstWouldAddEndPuncttrue
\mciteSetBstMidEndSepPunct{\mcitedefaultmidpunct}
{\mcitedefaultendpunct}{\mcitedefaultseppunct}\relax
\EndOfBibitem
\bibitem[Klicpera \latin{et~al.}(Apr. 26 -- May 1, 2020)Klicpera, Gro{\ss}, and
  G{\"{u}}nnemann]{Klicpera2020}
Klicpera,~J.; Gro{\ss},~J.; G{\"{u}}nnemann,~S. {Directional Message Passing
  for Molecular Graphs}. 8th International Conference on Learning
  Representations. Online, Apr. 26 -- May 1, 2020\relax
\mciteBstWouldAddEndPuncttrue
\mciteSetBstMidEndSepPunct{\mcitedefaultmidpunct}
{\mcitedefaultendpunct}{\mcitedefaultseppunct}\relax
\EndOfBibitem
\bibitem[Klicpera \latin{et~al.}(Dec. 12, 2020)Klicpera, Giri, Margraf, and
  G{\"{u}}nnemann]{Klicpera2020b}
Klicpera,~J.; Giri,~S.; Margraf,~J.~T.; G{\"{u}}nnemann,~S. Fast and
  Uncertainty-Aware Directional Message Passing for Non-Equilibrium Molecules.
  Machine Learning for Molecules Workshop at NeurIPS. Online, Dec. 12,
  2020\relax
\mciteBstWouldAddEndPuncttrue
\mciteSetBstMidEndSepPunct{\mcitedefaultmidpunct}
{\mcitedefaultendpunct}{\mcitedefaultseppunct}\relax
\EndOfBibitem
\bibitem[Ko \latin{et~al.}(2021)Ko, Finkler, Goedecker, and Behler]{Ko2021}
Ko,~T.~W.; Finkler,~J.~A.; Goedecker,~S.; Behler,~J. {A fourth-generation
  high-dimensional neural network potential with accurate electrostatics
  including non-local charge transfer}. \emph{Nat. Commun.} \textbf{2021},
  \emph{12}, 398\relax
\mciteBstWouldAddEndPuncttrue
\mciteSetBstMidEndSepPunct{\mcitedefaultmidpunct}
{\mcitedefaultendpunct}{\mcitedefaultseppunct}\relax
\EndOfBibitem
\bibitem[Batzner \latin{et~al.}(2022)Batzner, Musaelian, Sun, Geiger, Mailoa,
  Kornbluth, Molinari, Smidt, and Kozinsky]{Batzner2022}
Batzner,~S.; Musaelian,~A.; Sun,~L.; Geiger,~M.; Mailoa,~J.~P.; Kornbluth,~M.;
  Molinari,~N.; Smidt,~T.~E.; Kozinsky,~B. E (3)-equivariant graph neural
  networks for data-efficient and accurate interatomic potentials. \emph{Nat.
  Commun.} \textbf{2022}, \emph{13}, 2453\relax
\mciteBstWouldAddEndPuncttrue
\mciteSetBstMidEndSepPunct{\mcitedefaultmidpunct}
{\mcitedefaultendpunct}{\mcitedefaultseppunct}\relax
\EndOfBibitem
\bibitem[No{\'{e}} \latin{et~al.}(2020)No{\'{e}}, Tkatchenko, M{\"{u}}ller, and
  Clementi]{Noe2020}
No{\'{e}},~F.; Tkatchenko,~A.; M{\"{u}}ller,~K.~R.; Clementi,~C. {Machine
  Learning for Molecular Simulation}. \emph{Annu. Rev. Phys. Chem.}
  \textbf{2020}, \emph{71}, 361--390\relax
\mciteBstWouldAddEndPuncttrue
\mciteSetBstMidEndSepPunct{\mcitedefaultmidpunct}
{\mcitedefaultendpunct}{\mcitedefaultseppunct}\relax
\EndOfBibitem
\bibitem[Sch{\"{u}}tt \latin{et~al.}(2017)Sch{\"{u}}tt, Arbabzadah, Chmiela,
  M{\"{u}}ller, and Tkatchenko]{Schutt2017a}
Sch{\"{u}}tt,~K.~T.; Arbabzadah,~F.; Chmiela,~S.; M{\"{u}}ller,~K.~R.;
  Tkatchenko,~A. {Quantum-chemical insights from deep tensor neural networks}.
  \emph{Nat. Commun.} \textbf{2017}, \emph{8}, 13890\relax
\mciteBstWouldAddEndPuncttrue
\mciteSetBstMidEndSepPunct{\mcitedefaultmidpunct}
{\mcitedefaultendpunct}{\mcitedefaultseppunct}\relax
\EndOfBibitem
\bibitem[Faber \latin{et~al.}(2017)Faber, Hutchison, Huang, Gilmer, Schoenholz,
  Dahl, Vinyals, Kearnes, Riley, and {Von Lilienfeld}]{Faber2017}
Faber,~F.~A.; Hutchison,~L.; Huang,~B.; Gilmer,~J.; Schoenholz,~S.~S.;
  Dahl,~G.~E.; Vinyals,~O.; Kearnes,~S.; Riley,~P.~F.; {Von Lilienfeld},~O.~A.
  {Prediction Errors of Molecular Machine Learning Models Lower than Hybrid DFT
  Error}. \emph{J. Chem. Theory Comput.} \textbf{2017}, \emph{13},
  5255--5264\relax
\mciteBstWouldAddEndPuncttrue
\mciteSetBstMidEndSepPunct{\mcitedefaultmidpunct}
{\mcitedefaultendpunct}{\mcitedefaultseppunct}\relax
\EndOfBibitem
\bibitem[Stocker \latin{et~al.}(2022)Stocker, Gasteiger, Becker, G{\"u}nnemann,
  and Margraf]{Stocker2022}
Stocker,~S.; Gasteiger,~J.; Becker,~F.; G{\"u}nnemann,~S.; Margraf,~J.~T. How
  robust are modern graph neural network potentials in long and hot molecular
  dynamics simulations? \emph{Mach. Learn.: Sci. Technol.} \textbf{2022},
  \emph{3}, 045010\relax
\mciteBstWouldAddEndPuncttrue
\mciteSetBstMidEndSepPunct{\mcitedefaultmidpunct}
{\mcitedefaultendpunct}{\mcitedefaultseppunct}\relax
\EndOfBibitem
\bibitem[Wang \latin{et~al.}(2019)Wang, Olsson, Wehmeyer, P{\'{e}}rez, Charron,
  {De Fabritiis}, No{\'{e}}, and Clementi]{Wang2019}
Wang,~J.; Olsson,~S.; Wehmeyer,~C.; P{\'{e}}rez,~A.; Charron,~N.~E.; {De
  Fabritiis},~G.; No{\'{e}},~F.; Clementi,~C. {Machine Learning of
  Coarse-Grained Molecular Dynamics Force Fields}. \emph{ACS Cent. Sci.}
  \textbf{2019}, \emph{5}, 755--767\relax
\mciteBstWouldAddEndPuncttrue
\mciteSetBstMidEndSepPunct{\mcitedefaultmidpunct}
{\mcitedefaultendpunct}{\mcitedefaultseppunct}\relax
\EndOfBibitem
\bibitem[Thaler and Zavadlav(2021)Thaler, and Zavadlav]{Thaler_2021}
Thaler,~S.; Zavadlav,~J. Learning neural network potentials from experimental
  data via Differentiable Trajectory Reweighting. \emph{Nat. Commun.}
  \textbf{2021}, \emph{12}, 6884\relax
\mciteBstWouldAddEndPuncttrue
\mciteSetBstMidEndSepPunct{\mcitedefaultmidpunct}
{\mcitedefaultendpunct}{\mcitedefaultseppunct}\relax
\EndOfBibitem
\bibitem[van~der Oord \latin{et~al.}(2022)van~der Oord, Sachs, Kov{\'a}cs,
  Ortner, and Cs{\'a}nyi]{Van2022}
van~der Oord,~C.; Sachs,~M.; Kov{\'a}cs,~D.~P.; Ortner,~C.; Cs{\'a}nyi,~G.
  Hyperactive Learning (HAL) for Data-Driven Interatomic Potentials.
  \emph{arXiv:2210.04225} \textbf{2022}, \relax
\mciteBstWouldAddEndPunctfalse
\mciteSetBstMidEndSepPunct{\mcitedefaultmidpunct}
{}{\mcitedefaultseppunct}\relax
\EndOfBibitem
\bibitem[Gal \latin{et~al.}(2022)Gal, Koumoutsakos, Lanusse, Louppe, and
  Papadimitriou]{Gal2022}
Gal,~Y.; Koumoutsakos,~P.; Lanusse,~F.; Louppe,~G.; Papadimitriou,~C. Bayesian
  uncertainty quantification for machine-learned models in physics. \emph{Nat.
  Rev. Phys.} \textbf{2022}, \emph{4}, 573--577\relax
\mciteBstWouldAddEndPuncttrue
\mciteSetBstMidEndSepPunct{\mcitedefaultmidpunct}
{\mcitedefaultendpunct}{\mcitedefaultseppunct}\relax
\EndOfBibitem
\bibitem[Angelikopoulos \latin{et~al.}(2012)Angelikopoulos, Papadimitriou, and
  Koumoutsakos]{Angelikopoulos2012}
Angelikopoulos,~P.; Papadimitriou,~C.; Koumoutsakos,~P. Bayesian uncertainty
  quantification and propagation in molecular dynamics simulations: a high
  performance computing framework. \emph{J. Chem. Phys.} \textbf{2012},
  \emph{137}, 144103\relax
\mciteBstWouldAddEndPuncttrue
\mciteSetBstMidEndSepPunct{\mcitedefaultmidpunct}
{\mcitedefaultendpunct}{\mcitedefaultseppunct}\relax
\EndOfBibitem
\bibitem[Zavadlav \latin{et~al.}(2019)Zavadlav, Arampatzis, and
  Koumoutsakos]{Zavadlav2019}
Zavadlav,~J.; Arampatzis,~G.; Koumoutsakos,~P. Bayesian selection for
  coarse-grained models of liquid water. \emph{Sci. Rep.} \textbf{2019},
  \emph{9}, 99\relax
\mciteBstWouldAddEndPuncttrue
\mciteSetBstMidEndSepPunct{\mcitedefaultmidpunct}
{\mcitedefaultendpunct}{\mcitedefaultseppunct}\relax
\EndOfBibitem
\bibitem[Wan \latin{et~al.}(2021)Wan, Sinclair, and Coveney]{Wan2021}
Wan,~S.; Sinclair,~R.~C.; Coveney,~P.~V. Uncertainty quantification in
  classical molecular dynamics. \emph{Philos. Trans. Royal Soc. A}
  \textbf{2021}, \emph{379}, 20200082\relax
\mciteBstWouldAddEndPuncttrue
\mciteSetBstMidEndSepPunct{\mcitedefaultmidpunct}
{\mcitedefaultendpunct}{\mcitedefaultseppunct}\relax
\EndOfBibitem
\bibitem[Smith \latin{et~al.}(2018)Smith, Nebgen, Lubbers, Isayev, and
  Roitberg]{Smith2018}
Smith,~J.~S.; Nebgen,~B.; Lubbers,~N.; Isayev,~O.; Roitberg,~A.~E. Less is
  more: Sampling chemical space with active learning. \emph{J. Chem. Phys.}
  \textbf{2018}, \emph{148}, 241733\relax
\mciteBstWouldAddEndPuncttrue
\mciteSetBstMidEndSepPunct{\mcitedefaultmidpunct}
{\mcitedefaultendpunct}{\mcitedefaultseppunct}\relax
\EndOfBibitem
\bibitem[Zhang \latin{et~al.}(2019)Zhang, Lin, Wang, Car, and
  Weinan]{Zhang2019}
Zhang,~L.; Lin,~D.-Y.; Wang,~H.; Car,~R.; Weinan,~E. Active learning of
  uniformly accurate interatomic potentials for materials simulation.
  \emph{Phys. Rev. Mater.} \textbf{2019}, \emph{3}, 023804\relax
\mciteBstWouldAddEndPuncttrue
\mciteSetBstMidEndSepPunct{\mcitedefaultmidpunct}
{\mcitedefaultendpunct}{\mcitedefaultseppunct}\relax
\EndOfBibitem
\bibitem[Loeffler \latin{et~al.}(2020)Loeffler, Patra, Chan, and
  Sankaranarayanan]{Loeffler2020}
Loeffler,~T.~D.; Patra,~T.~K.; Chan,~H.; Sankaranarayanan,~S.~K. Active
  learning a coarse-grained neural network model for bulk water from sparse
  training data. \emph{Mol. Syst. Des. Eng.} \textbf{2020}, \emph{5},
  902--910\relax
\mciteBstWouldAddEndPuncttrue
\mciteSetBstMidEndSepPunct{\mcitedefaultmidpunct}
{\mcitedefaultendpunct}{\mcitedefaultseppunct}\relax
\EndOfBibitem
\bibitem[Smith \latin{et~al.}(2021)Smith, Nebgen, Mathew, Chen, Lubbers,
  Burakovsky, Tretiak, Nam, Germann, Fensin, \latin{et~al.} others]{Smith2021}
Smith,~J.~S.; Nebgen,~B.; Mathew,~N.; Chen,~J.; Lubbers,~N.; Burakovsky,~L.;
  Tretiak,~S.; Nam,~H.~A.; Germann,~T.; Fensin,~S., \latin{et~al.}  Automated
  discovery of a robust interatomic potential for aluminum. \emph{Nat. Commun.}
  \textbf{2021}, \emph{12}, 1257\relax
\mciteBstWouldAddEndPuncttrue
\mciteSetBstMidEndSepPunct{\mcitedefaultmidpunct}
{\mcitedefaultendpunct}{\mcitedefaultseppunct}\relax
\EndOfBibitem
\bibitem[Xie \latin{et~al.}(Mar. 14--18, 2021)Xie, Rupp, and Hennig]{Xie2021}
Xie,~S.~R.; Rupp,~M.; Hennig,~R.~G. Ultra-fast Force Fields (UF3) framework for
  machine-learning interatomic potentials. American Physical Society March
  Meeting. Chicago, IL, USA, Mar. 14--18, 2021\relax
\mciteBstWouldAddEndPuncttrue
\mciteSetBstMidEndSepPunct{\mcitedefaultmidpunct}
{\mcitedefaultendpunct}{\mcitedefaultseppunct}\relax
\EndOfBibitem
\bibitem[Imbalzano \latin{et~al.}(2021)Imbalzano, Zhuang, Kapil, Rossi, Engel,
  Grasselli, and Ceriotti]{Imbalzano2021}
Imbalzano,~G.; Zhuang,~Y.; Kapil,~V.; Rossi,~K.; Engel,~E.~A.; Grasselli,~F.;
  Ceriotti,~M. Uncertainty estimation for molecular dynamics and sampling.
  \emph{J. Chem. Phys.} \textbf{2021}, \emph{154}, 074102\relax
\mciteBstWouldAddEndPuncttrue
\mciteSetBstMidEndSepPunct{\mcitedefaultmidpunct}
{\mcitedefaultendpunct}{\mcitedefaultseppunct}\relax
\EndOfBibitem
\bibitem[Duane \latin{et~al.}(1987)Duane, Kennedy, Pendleton, and
  Roweth]{Duane1987}
Duane,~S.; Kennedy,~A.~D.; Pendleton,~B.~J.; Roweth,~D. Hybrid Monte Carlo.
  \emph{Phys. Lett. B} \textbf{1987}, \emph{195}, 216--222\relax
\mciteBstWouldAddEndPuncttrue
\mciteSetBstMidEndSepPunct{\mcitedefaultmidpunct}
{\mcitedefaultendpunct}{\mcitedefaultseppunct}\relax
\EndOfBibitem
\bibitem[Neal(2011)]{Neal2011}
Neal,~R.~M. In \emph{Handbook of Markov Chain Monte Carlo}, 1st ed.;
  Brooks,~S., Gelman,~A., Jones,~G.~L., Meng,~X.-L., Eds.; Chapman and
  Hall/CRC: New York, USA, 2011; Chapter MCMC using Hamiltonian Dynamics, pp
  139--188\relax
\mciteBstWouldAddEndPuncttrue
\mciteSetBstMidEndSepPunct{\mcitedefaultmidpunct}
{\mcitedefaultendpunct}{\mcitedefaultseppunct}\relax
\EndOfBibitem
\bibitem[Kahle and Zipoli(2022)Kahle, and Zipoli]{Kahle2022}
Kahle,~L.; Zipoli,~F. Quality of uncertainty estimates from neural network
  potential ensembles. \emph{Phys. Rev. E} \textbf{2022}, \emph{105},
  015311\relax
\mciteBstWouldAddEndPuncttrue
\mciteSetBstMidEndSepPunct{\mcitedefaultmidpunct}
{\mcitedefaultendpunct}{\mcitedefaultseppunct}\relax
\EndOfBibitem
\bibitem[Welling and Teh(Jun. 28 -- Jul. 2, 2011)Welling, and Teh]{Welling2011}
Welling,~M.; Teh,~Y.~W. Bayesian learning via stochastic gradient Langevin
  dynamics. Proceedings of the 28th International Conference on Machine
  Learning. Bellevue, WA, USA, Jun. 28 -- Jul. 2, 2011; pp 681--688\relax
\mciteBstWouldAddEndPuncttrue
\mciteSetBstMidEndSepPunct{\mcitedefaultmidpunct}
{\mcitedefaultendpunct}{\mcitedefaultseppunct}\relax
\EndOfBibitem
\bibitem[Chen \latin{et~al.}(Jun. 21 --26, 2014)Chen, Fox, and
  Guestrin]{Chen2014}
Chen,~T.; Fox,~E.; Guestrin,~C. Stochastic gradient Hamiltonian Monte Carlo.
  Proceedings of the 31st International Conference on Machine Learning.
  Beijing, China, Jun. 21 --26, 2014; pp 1683--1691\relax
\mciteBstWouldAddEndPuncttrue
\mciteSetBstMidEndSepPunct{\mcitedefaultmidpunct}
{\mcitedefaultendpunct}{\mcitedefaultseppunct}\relax
\EndOfBibitem
\bibitem[Li \latin{et~al.}(February 12--17, 2016)Li, Chen, Carlson, and
  Carin]{Li2016}
Li,~C.; Chen,~C.; Carlson,~D.~E.; Carin,~L. Preconditioned Stochastic Gradient
  Langevin Dynamics for Deep Neural Networks. Proceedings of the Thirtieth
  {AAAI} Conference on Artificial Intelligence. Phoenix, AZ, {USA}, February
  12--17, 2016; pp 1788--1794\relax
\mciteBstWouldAddEndPuncttrue
\mciteSetBstMidEndSepPunct{\mcitedefaultmidpunct}
{\mcitedefaultendpunct}{\mcitedefaultseppunct}\relax
\EndOfBibitem
\bibitem[Nemeth and Fearnhead(2021)Nemeth, and Fearnhead]{Nemeth2021}
Nemeth,~C.; Fearnhead,~P. Stochastic gradient Markov chain Monte Carlo.
  \emph{J. Am. Stat. Assoc.} \textbf{2021}, \emph{116}, 433--450\relax
\mciteBstWouldAddEndPuncttrue
\mciteSetBstMidEndSepPunct{\mcitedefaultmidpunct}
{\mcitedefaultendpunct}{\mcitedefaultseppunct}\relax
\EndOfBibitem
\bibitem[Lamb and Paige(Dec. 12, 2020)Lamb, and Paige]{Lamb2020}
Lamb,~G.; Paige,~B. Bayesian Graph Neural Networks for Molecular Property
  Prediction. Machine Learning for Molecules Workshop at NeurIPS. Online, Dec.
  12, 2020\relax
\mciteBstWouldAddEndPuncttrue
\mciteSetBstMidEndSepPunct{\mcitedefaultmidpunct}
{\mcitedefaultendpunct}{\mcitedefaultseppunct}\relax
\EndOfBibitem
\bibitem[Graves(Dec. 12 -- 14, 2011)]{Graves2011}
Graves,~A. Practical Variational Inference for Neural Networks. Advances in
  Neural Information Processing Systems. Granada, Spain, Dec. 12 -- 14,
  2011\relax
\mciteBstWouldAddEndPuncttrue
\mciteSetBstMidEndSepPunct{\mcitedefaultmidpunct}
{\mcitedefaultendpunct}{\mcitedefaultseppunct}\relax
\EndOfBibitem
\bibitem[Hoffman \latin{et~al.}(2013)Hoffman, Blei, Wang, and
  Paisley]{Hoffman2013}
Hoffman,~M.~D.; Blei,~D.~M.; Wang,~C.; Paisley,~J. Stochastic Variational
  Inference. \emph{J. Mach. Learn. Res.} \textbf{2013}, \emph{14},
  1303--1347\relax
\mciteBstWouldAddEndPuncttrue
\mciteSetBstMidEndSepPunct{\mcitedefaultmidpunct}
{\mcitedefaultendpunct}{\mcitedefaultseppunct}\relax
\EndOfBibitem
\bibitem[Hansen and Salamon(1990)Hansen, and Salamon]{Hansen90}
Hansen,~L.; Salamon,~P. Neural Network Ensembles. \emph{IEEE Trans. Pattern
  Anal. Machine Intell.} \textbf{1990}, \emph{12}, 993--1001\relax
\mciteBstWouldAddEndPuncttrue
\mciteSetBstMidEndSepPunct{\mcitedefaultmidpunct}
{\mcitedefaultendpunct}{\mcitedefaultseppunct}\relax
\EndOfBibitem
\bibitem[Lakshminarayanan \latin{et~al.}(Dec. 4--9, 2017)Lakshminarayanan,
  Pritzel, and Blundell]{Lakshminarayanan2017}
Lakshminarayanan,~B.; Pritzel,~A.; Blundell,~C. Simple and Scalable Predictive
  Uncertainty Estimation using Deep Ensembles. Advances in Neural Information
  Processing Systems. Long Beach, CA, USA, Dec. 4--9, 2017\relax
\mciteBstWouldAddEndPuncttrue
\mciteSetBstMidEndSepPunct{\mcitedefaultmidpunct}
{\mcitedefaultendpunct}{\mcitedefaultseppunct}\relax
\EndOfBibitem
\bibitem[Ovadia \latin{et~al.}(Dec. 8 -- 14, 2019)Ovadia, Fertig, Ren, Nado,
  Sculley, Nowozin, Dillon, Lakshminarayanan, and Snoek]{Ovadia2019}
Ovadia,~Y.; Fertig,~E.; Ren,~J.; Nado,~Z.; Sculley,~D.; Nowozin,~S.;
  Dillon,~J.; Lakshminarayanan,~B.; Snoek,~J. Can you trust your model's
  uncertainty? Evaluating predictive uncertainty under dataset shift. Advances
  in Neural Information Processing Systems. Vancouver, BC, Canada, Dec. 8 --
  14, 2019\relax
\mciteBstWouldAddEndPuncttrue
\mciteSetBstMidEndSepPunct{\mcitedefaultmidpunct}
{\mcitedefaultendpunct}{\mcitedefaultseppunct}\relax
\EndOfBibitem
\bibitem[Wenzel \latin{et~al.}(Jul. 13 -- 18, 2020)Wenzel, Roth, Veeling,
  Swik{a}tkowski, Tran, Mandt, Snoek, Salimans, Jenatton, and
  Nowozin]{Wenzel2020}
Wenzel,~F.; Roth,~K.; Veeling,~B.~S.; Swik{a}tkowski,~J.; Tran,~L.; Mandt,~S.;
  Snoek,~J.; Salimans,~T.; Jenatton,~R.; Nowozin,~S. How Good is the Bayes
  Posterior in Deep Neural Networks Really? Proceedings of the 37th
  International Conference on Machine Learning. Online, Jul. 13 -- 18, 2020; pp
  10248--10259\relax
\mciteBstWouldAddEndPuncttrue
\mciteSetBstMidEndSepPunct{\mcitedefaultmidpunct}
{\mcitedefaultendpunct}{\mcitedefaultseppunct}\relax
\EndOfBibitem
\bibitem[Wen and Tadmor(2020)Wen, and Tadmor]{Wen2020}
Wen,~M.; Tadmor,~E.~B. Uncertainty quantification in molecular simulations with
  dropout neural network potentials. \emph{Npj Comput. Mater.} \textbf{2020},
  \emph{6}, 124\relax
\mciteBstWouldAddEndPuncttrue
\mciteSetBstMidEndSepPunct{\mcitedefaultmidpunct}
{\mcitedefaultendpunct}{\mcitedefaultseppunct}\relax
\EndOfBibitem
\bibitem[Zhu \latin{et~al.}(2022)Zhu, Batzner, Musaelian, and
  Kozinsky]{Zhu2022}
Zhu,~A.; Batzner,~S.; Musaelian,~A.; Kozinsky,~B. Fast Uncertainty Estimates in
  Deep Learning Interatomic Potentials. \emph{arXiv preprint arXiv:2211.09866}
  \textbf{2022}, \relax
\mciteBstWouldAddEndPunctfalse
\mciteSetBstMidEndSepPunct{\mcitedefaultmidpunct}
{}{\mcitedefaultseppunct}\relax
\EndOfBibitem
\bibitem[Gustafsson \latin{et~al.}(Jun. 14 -- 19, 2020)Gustafsson, Danelljan,
  and Schon]{Gustafsson2020}
Gustafsson,~F.~K.; Danelljan,~M.; Schon,~T.~B. Evaluating Scalable Bayesian
  Deep Learning Methods for Robust Computer Vision. Proceedings of the IEEE/CVF
  Conference on Computer Vision and Pattern Recognition (CVPR) Workshops.
  Seattle, WA, USA, Jun. 14 -- 19, 2020; pp 1289--1298\relax
\mciteBstWouldAddEndPuncttrue
\mciteSetBstMidEndSepPunct{\mcitedefaultmidpunct}
{\mcitedefaultendpunct}{\mcitedefaultseppunct}\relax
\EndOfBibitem
\bibitem[H{\"u}llermeier and Waegeman(2021)H{\"u}llermeier, and
  Waegeman]{Hullermeier2021}
H{\"u}llermeier,~E.; Waegeman,~W. Aleatoric and epistemic uncertainty in
  machine learning: An introduction to concepts and methods. \emph{Mach.
  Learn.} \textbf{2021}, \emph{110}, 457--506\relax
\mciteBstWouldAddEndPuncttrue
\mciteSetBstMidEndSepPunct{\mcitedefaultmidpunct}
{\mcitedefaultendpunct}{\mcitedefaultseppunct}\relax
\EndOfBibitem
\bibitem[Wilson and Izmailov(Dec. 6--12, 2020)Wilson, and Izmailov]{Wilson2020}
Wilson,~A.~G.; Izmailov,~P. Bayesian Deep Learning and a Probabilistic
  Perspective of Generalization. Advances in Neural Information Processing
  Systems. Online, Dec. 6--12, 2020\relax
\mciteBstWouldAddEndPuncttrue
\mciteSetBstMidEndSepPunct{\mcitedefaultmidpunct}
{\mcitedefaultendpunct}{\mcitedefaultseppunct}\relax
\EndOfBibitem
\bibitem[Izmailov \latin{et~al.}(Jul. 18--44, 2021)Izmailov, Vikram, Hoffman,
  and Wilson]{Izmailov2021}
Izmailov,~P.; Vikram,~S.; Hoffman,~M.~D.; Wilson,~A. G.~G. What Are Bayesian
  Neural Network Posteriors Really Like? Proceedings of the 38th International
  Conference on Machine Learning. Online, Jul. 18--44, 2021; pp
  4629--4640\relax
\mciteBstWouldAddEndPuncttrue
\mciteSetBstMidEndSepPunct{\mcitedefaultmidpunct}
{\mcitedefaultendpunct}{\mcitedefaultseppunct}\relax
\EndOfBibitem
\bibitem[Hastings(1970)]{Hastings1970}
Hastings,~W.~K. Monte Carlo sampling methods using Markov chains and their
  applications. \emph{Biometrika} \textbf{1970}, \emph{57}, 97--109\relax
\mciteBstWouldAddEndPuncttrue
\mciteSetBstMidEndSepPunct{\mcitedefaultmidpunct}
{\mcitedefaultendpunct}{\mcitedefaultseppunct}\relax
\EndOfBibitem
\bibitem[Li and Br{\"{u}}schweiler(2011)Li, and Br{\"{u}}schweiler]{Li2011}
Li,~D.~W.; Br{\"{u}}schweiler,~R. {Iterative Optimization of Molecular
  Mechanics Force Fields from NMR Data of Full-Length Proteins}. \emph{J. Chem.
  Theory Comput.} \textbf{2011}, \emph{7}, 1773--1782\relax
\mciteBstWouldAddEndPuncttrue
\mciteSetBstMidEndSepPunct{\mcitedefaultmidpunct}
{\mcitedefaultendpunct}{\mcitedefaultseppunct}\relax
\EndOfBibitem
\bibitem[Teh \latin{et~al.}(2016)Teh, Thiery, and Vollmer]{Teh2016}
Teh,~Y.~W.; Thiery,~A.~H.; Vollmer,~S.~J. Consistency and Fluctuations For
  Stochastic Gradient Langevin Dynamics. \emph{J. Mach. Learn. Res.}
  \textbf{2016}, \emph{17}, 1--33\relax
\mciteBstWouldAddEndPuncttrue
\mciteSetBstMidEndSepPunct{\mcitedefaultmidpunct}
{\mcitedefaultendpunct}{\mcitedefaultseppunct}\relax
\EndOfBibitem
\bibitem[Tieleman and Hinton(2012)Tieleman, and Hinton]{Tieleman2012}
Tieleman,~T.; Hinton,~G. Lecture 6.5-rmsprop: Divide the gradient by a running
  average of its recent magnitude. \emph{COURSERA: Neural networks for machine
  learning} \textbf{2012}, \emph{4}, 26--31\relax
\mciteBstWouldAddEndPuncttrue
\mciteSetBstMidEndSepPunct{\mcitedefaultmidpunct}
{\mcitedefaultendpunct}{\mcitedefaultseppunct}\relax
\EndOfBibitem
\bibitem[Dauphin \latin{et~al.}(Dec. 8--13, 2014)Dauphin, Pascanu, Gulcehre,
  Cho, Ganguli, and Bengio]{Dauphin2014}
Dauphin,~Y.~N.; Pascanu,~R.; Gulcehre,~C.; Cho,~K.; Ganguli,~S.; Bengio,~Y.
  Identifying and attacking the saddle point problem in high-dimensional
  non-convex optimization. Advances in Neural Information Processing Systems.
  Montreal, QC, Canada, Dec. 8--13, 2014\relax
\mciteBstWouldAddEndPuncttrue
\mciteSetBstMidEndSepPunct{\mcitedefaultmidpunct}
{\mcitedefaultendpunct}{\mcitedefaultseppunct}\relax
\EndOfBibitem
\bibitem[Thaler \latin{et~al.}(2020)Thaler, Fuchs, Cukarska, and
  Zavadlav]{Thaler2022c}
Thaler,~S.; Fuchs,~P.; Cukarska,~A.; Zavadlav,~J. jax-sgmc: Modular Stochastic
  Gradient {MCMC} for {JAX}. 2020;
  \url{https://github.com/tummfm/jax-sgmc}\relax
\mciteBstWouldAddEndPuncttrue
\mciteSetBstMidEndSepPunct{\mcitedefaultmidpunct}
{\mcitedefaultendpunct}{\mcitedefaultseppunct}\relax
\EndOfBibitem
\bibitem[Fort \latin{et~al.}(2019)Fort, Hu, and Lakshminarayanan]{Fort2019}
Fort,~S.; Hu,~H.; Lakshminarayanan,~B. Deep ensembles: A loss landscape
  perspective. \emph{arXiv preprint arXiv:1912.02757} \textbf{2019}, \relax
\mciteBstWouldAddEndPunctfalse
\mciteSetBstMidEndSepPunct{\mcitedefaultmidpunct}
{}{\mcitedefaultseppunct}\relax
\EndOfBibitem
\bibitem[Smith \latin{et~al.}(2020)Smith, Zubatyuk, Nebgen, Lubbers, Barros,
  Roitberg, Isayev, and Tretiak]{Smith2020}
Smith,~J.~S.; Zubatyuk,~R.; Nebgen,~B.; Lubbers,~N.; Barros,~K.;
  Roitberg,~A.~E.; Isayev,~O.; Tretiak,~S. The ANI-1ccx and ANI-1x data sets,
  coupled-cluster and density functional theory properties for molecules.
  \emph{Sci. Data} \textbf{2020}, \emph{7}, 134\relax
\mciteBstWouldAddEndPuncttrue
\mciteSetBstMidEndSepPunct{\mcitedefaultmidpunct}
{\mcitedefaultendpunct}{\mcitedefaultseppunct}\relax
\EndOfBibitem
\bibitem[Izvekov and Voth(2005)Izvekov, and Voth]{Izvekov2005}
Izvekov,~S.; Voth,~G.~A. Multiscale coarse graining of liquid-state systems.
  \emph{J. Chem. Phys.} \textbf{2005}, \emph{123}, 134105\relax
\mciteBstWouldAddEndPuncttrue
\mciteSetBstMidEndSepPunct{\mcitedefaultmidpunct}
{\mcitedefaultendpunct}{\mcitedefaultseppunct}\relax
\EndOfBibitem
\bibitem[Noid \latin{et~al.}(2008)Noid, Chu, Ayton, Krishna, Izvekov, Voth,
  Das, and Andersen]{Noid2008}
Noid,~W.~G.; Chu,~J.-W.; Ayton,~G.~S.; Krishna,~V.; Izvekov,~S.; Voth,~G.~A.;
  Das,~A.; Andersen,~H.~C. The multiscale coarse-graining method. I. A rigorous
  bridge between atomistic and coarse-grained models. \emph{J. Chem. Phys.}
  \textbf{2008}, \emph{128}, 244114\relax
\mciteBstWouldAddEndPuncttrue
\mciteSetBstMidEndSepPunct{\mcitedefaultmidpunct}
{\mcitedefaultendpunct}{\mcitedefaultseppunct}\relax
\EndOfBibitem
\bibitem[Noid \latin{et~al.}(2008)Noid, Liu, Wang, Chu, Ayton, Izvekov,
  Andersen, and Voth]{Noid2008b}
Noid,~W.; Liu,~P.; Wang,~Y.; Chu,~J.-W.; Ayton,~G.~S.; Izvekov,~S.;
  Andersen,~H.~C.; Voth,~G.~A. The multiscale coarse-graining method. II.
  Numerical implementation for coarse-grained molecular models. \emph{J. Chem.
  Phys.} \textbf{2008}, \emph{128}, 244115\relax
\mciteBstWouldAddEndPuncttrue
\mciteSetBstMidEndSepPunct{\mcitedefaultmidpunct}
{\mcitedefaultendpunct}{\mcitedefaultseppunct}\relax
\EndOfBibitem
\bibitem[Wang \latin{et~al.}(2009)Wang, Junghans, and Kremer]{Wang2009}
Wang,~H.; Junghans,~C.; Kremer,~K. {Comparative atomistic and coarse-grained
  study of water: What do we lose by coarse-graining?} \emph{Eur. Phys. J. E}
  \textbf{2009}, \emph{28}, 221--229\relax
\mciteBstWouldAddEndPuncttrue
\mciteSetBstMidEndSepPunct{\mcitedefaultmidpunct}
{\mcitedefaultendpunct}{\mcitedefaultseppunct}\relax
\EndOfBibitem
\bibitem[Fu \latin{et~al.}(Dec. 2, 2022)Fu, Wu, Wang, Xie, Keten,
  Gomez-Bombarelli, and Jaakkola]{Fu2022}
Fu,~X.; Wu,~Z.; Wang,~W.; Xie,~T.; Keten,~S.; Gomez-Bombarelli,~R.;
  Jaakkola,~T. Forces are not Enough: Benchmark and Critical Evaluation for
  Machine Learning Force Fields with Molecular Simulations. AI for Science:
  Progress and Promises Workshop at NeurIPS. New Orleans, LA, USA, Dec. 2,
  2022\relax
\mciteBstWouldAddEndPuncttrue
\mciteSetBstMidEndSepPunct{\mcitedefaultmidpunct}
{\mcitedefaultendpunct}{\mcitedefaultseppunct}\relax
\EndOfBibitem
\bibitem[Das and Andersen(2009)Das, and Andersen]{Das2009}
Das,~A.; Andersen,~H.~C. The multiscale coarse-graining method. III. A test of
  pairwise additivity of the coarse-grained potential and of new basis
  functions for the variational calculation. \emph{J. Chem. Phys.}
  \textbf{2009}, \emph{131}, 034102\relax
\mciteBstWouldAddEndPuncttrue
\mciteSetBstMidEndSepPunct{\mcitedefaultmidpunct}
{\mcitedefaultendpunct}{\mcitedefaultseppunct}\relax
\EndOfBibitem
\bibitem[Husic \latin{et~al.}(2020)Husic, Charron, Lemm, Wang, P{\'{e}}rez,
  Kr{\"{a}}mer, Chen, Olsson, de~Fabritiis, No{\'{e}}, and Clementi]{Husic2020}
Husic,~B.~E.; Charron,~N.~E.; Lemm,~D.; Wang,~J.; P{\'{e}}rez,~A.;
  Kr{\"{a}}mer,~A.; Chen,~Y.; Olsson,~S.; de~Fabritiis,~G.; No{\'{e}},~F.;
  Clementi,~C. Coarse graining molecular dynamics with graph neural networks.
  \emph{J. Chem. Phys.} \textbf{2020}, \emph{153}, 194101\relax
\mciteBstWouldAddEndPuncttrue
\mciteSetBstMidEndSepPunct{\mcitedefaultmidpunct}
{\mcitedefaultendpunct}{\mcitedefaultseppunct}\relax
\EndOfBibitem
\bibitem[Thaler \latin{et~al.}(2022)Thaler, Stupp, and Zavadlav]{Thaler2022}
Thaler,~S.; Stupp,~M.; Zavadlav,~J. Deep coarse-grained potentials via relative
  entropy minimization. \emph{J. Chem. Phys.} \textbf{2022}, \emph{157},
  244103\relax
\mciteBstWouldAddEndPuncttrue
\mciteSetBstMidEndSepPunct{\mcitedefaultmidpunct}
{\mcitedefaultendpunct}{\mcitedefaultseppunct}\relax
\EndOfBibitem
\bibitem[Hoffman and Gelman(2014)Hoffman, and Gelman]{Hoffman2014}
Hoffman,~M.~D.; Gelman,~A. The No-U-Turn sampler: adaptively setting path
  lengths in Hamiltonian Monte Carlo. \emph{J. Mach. Learn. Res.}
  \textbf{2014}, \emph{15}, 1593--1623\relax
\mciteBstWouldAddEndPuncttrue
\mciteSetBstMidEndSepPunct{\mcitedefaultmidpunct}
{\mcitedefaultendpunct}{\mcitedefaultseppunct}\relax
\EndOfBibitem
\bibitem[Gelman \latin{et~al.}(2015)Gelman, Lee, and Guo]{Gelman2015}
Gelman,~A.; Lee,~D.; Guo,~J. Stan: A probabilistic programming language for
  Bayesian inference and optimization. \emph{J. Educ. Behav. Stat.}
  \textbf{2015}, \emph{40}, 530--543\relax
\mciteBstWouldAddEndPuncttrue
\mciteSetBstMidEndSepPunct{\mcitedefaultmidpunct}
{\mcitedefaultendpunct}{\mcitedefaultseppunct}\relax
\EndOfBibitem
\bibitem[Lao and Louf(2020)Lao, and Louf]{Blackjax2020}
Lao,~J.; Louf,~R. {B}lackjax: A sampling library for {JAX}. 2020;
  \url{http://github.com/blackjax-devs/blackjax}\relax
\mciteBstWouldAddEndPuncttrue
\mciteSetBstMidEndSepPunct{\mcitedefaultmidpunct}
{\mcitedefaultendpunct}{\mcitedefaultseppunct}\relax
\EndOfBibitem
\bibitem[Tran \latin{et~al.}(2020)Tran, Neiswanger, Yoon, Zhang, Xing, and
  Ulissi]{Tran2020}
Tran,~K.; Neiswanger,~W.; Yoon,~J.; Zhang,~Q.; Xing,~E.; Ulissi,~Z.~W. Methods
  for comparing uncertainty quantifications for material property predictions.
  \emph{Mach. Learn. Sci. Technol.} \textbf{2020}, \emph{1}, 025006\relax
\mciteBstWouldAddEndPuncttrue
\mciteSetBstMidEndSepPunct{\mcitedefaultmidpunct}
{\mcitedefaultendpunct}{\mcitedefaultseppunct}\relax
\EndOfBibitem
\bibitem[Chung \latin{et~al.}(2021)Chung, Char, Guo, Schneider, and
  Neiswanger]{Chung2021}
Chung,~Y.; Char,~I.; Guo,~H.; Schneider,~J.; Neiswanger,~W. Uncertainty
  Toolbox: an Open-Source Library for Assessing, Visualizing, and Improving
  Uncertainty Quantification. \emph{arXiv preprint arXiv:2109.10254}
  \textbf{2021}, \relax
\mciteBstWouldAddEndPunctfalse
\mciteSetBstMidEndSepPunct{\mcitedefaultmidpunct}
{}{\mcitedefaultseppunct}\relax
\EndOfBibitem
\bibitem[Hirschfeld \latin{et~al.}(2020)Hirschfeld, Swanson, Yang, Barzilay,
  and Coley]{Hirschfeld2020}
Hirschfeld,~L.; Swanson,~K.; Yang,~K.; Barzilay,~R.; Coley,~C.~W. Uncertainty
  quantification using neural networks for molecular property prediction.
  \emph{J. Chem. Inf. Model.} \textbf{2020}, \emph{60}, 3770--3780\relax
\mciteBstWouldAddEndPuncttrue
\mciteSetBstMidEndSepPunct{\mcitedefaultmidpunct}
{\mcitedefaultendpunct}{\mcitedefaultseppunct}\relax
\EndOfBibitem
\bibitem[Abascal and Vega(2005)Abascal, and Vega]{Abascal2005}
Abascal,~J. L.~F.; Vega,~C. A general purpose model for the condensed phases of
  water: TIP4P/2005. \emph{J. Chem. Phys.} \textbf{2005}, \emph{123},
  234505\relax
\mciteBstWouldAddEndPuncttrue
\mciteSetBstMidEndSepPunct{\mcitedefaultmidpunct}
{\mcitedefaultendpunct}{\mcitedefaultseppunct}\relax
\EndOfBibitem
\bibitem[Berendsen \latin{et~al.}(Apr. 13--16, 1981)Berendsen, Postma, van
  Gunsteren, and Hermans]{Berendsen1981}
Berendsen,~H. J.~C.; Postma,~J. P.~M.; van Gunsteren,~W.~F.; Hermans,~J.
  Interaction models for water in relation to protein hydration. Intermolecular
  forces: Proceedings of the Fourteenth Jerusalem Symposium on Quantum
  Chemistry and Biochemistry. Jerusalem, Israel, Apr. 13--16, 1981; pp
  331--342\relax
\mciteBstWouldAddEndPuncttrue
\mciteSetBstMidEndSepPunct{\mcitedefaultmidpunct}
{\mcitedefaultendpunct}{\mcitedefaultseppunct}\relax
\EndOfBibitem
\bibitem[Chaimovich and Shell(2009)Chaimovich, and Shell]{Chaimovich2009}
Chaimovich,~A.; Shell,~M.~S. Anomalous waterlike behavior in
  spherically-symmetric water models optimized with the relative entropy.
  \emph{Phys. Chem. Chem. Phys.} \textbf{2009}, \emph{11}, 1901--1915\relax
\mciteBstWouldAddEndPuncttrue
\mciteSetBstMidEndSepPunct{\mcitedefaultmidpunct}
{\mcitedefaultendpunct}{\mcitedefaultseppunct}\relax
\EndOfBibitem
\bibitem[Potestio \latin{et~al.}(2014)Potestio, Peter, and
  Kremer]{Potestio2014}
Potestio,~R.; Peter,~C.; Kremer,~K. Computer Simulations of Soft matter:
  Linking the Scales. \emph{Entropy} \textbf{2014}, \emph{16}, 4199--4245\relax
\mciteBstWouldAddEndPuncttrue
\mciteSetBstMidEndSepPunct{\mcitedefaultmidpunct}
{\mcitedefaultendpunct}{\mcitedefaultseppunct}\relax
\EndOfBibitem
\bibitem[Kingma and Ba(May 7-9, 2015)Kingma, and Ba]{Kingma2015}
Kingma,~D.~P.; Ba,~J.~L. {Adam: A Method for Stochastic Optimization}. 3rd
  International Conference on Learning Representations, {ICLR}. San Diego, CA,
  USA, May 7-9, 2015\relax
\mciteBstWouldAddEndPuncttrue
\mciteSetBstMidEndSepPunct{\mcitedefaultmidpunct}
{\mcitedefaultendpunct}{\mcitedefaultseppunct}\relax
\EndOfBibitem
\bibitem[Thaler and Zavadlav(Jul. 27 --29, 2022)Thaler, and
  Zavadlav]{Thaler2022b}
Thaler,~S.; Zavadlav,~J. Uncertainty Quantification for Molecular Models via
  Stochastic Gradient MCMC. 10th Vienna Conference on Mathematical Modelling.
  Vienna, Austria, Jul. 27 --29, 2022; pp 19--20\relax
\mciteBstWouldAddEndPuncttrue
\mciteSetBstMidEndSepPunct{\mcitedefaultmidpunct}
{\mcitedefaultendpunct}{\mcitedefaultseppunct}\relax
\EndOfBibitem
\bibitem[Pettitt and Karplus(1985)Pettitt, and Karplus]{Pettitt1985}
Pettitt,~B.~M.; Karplus,~M. The potential of mean force surface for the alanine
  dipeptide in aqueous solution: a theoretical approach. \emph{Chem. Phys.
  Lett.} \textbf{1985}, \emph{121}, 194--201\relax
\mciteBstWouldAddEndPuncttrue
\mciteSetBstMidEndSepPunct{\mcitedefaultmidpunct}
{\mcitedefaultendpunct}{\mcitedefaultseppunct}\relax
\EndOfBibitem
\bibitem[Tobias and Brooks~III(1992)Tobias, and Brooks~III]{Tobias1992}
Tobias,~D.~J.; Brooks~III,~C.~L. Conformational equilibrium in the alanine
  dipeptide in the gas phase and aqueous solution: A comparison of theoretical
  results. \emph{J. Phys. Chem.} \textbf{1992}, \emph{96}, 3864--3870\relax
\mciteBstWouldAddEndPuncttrue
\mciteSetBstMidEndSepPunct{\mcitedefaultmidpunct}
{\mcitedefaultendpunct}{\mcitedefaultseppunct}\relax
\EndOfBibitem
\bibitem[Shell(2008)]{Shell2008}
Shell,~M.~S. The relative entropy is fundamental to multiscale and inverse
  thermodynamic problems. \emph{J. Chem. Phys.} \textbf{2008}, \emph{129},
  144108\relax
\mciteBstWouldAddEndPuncttrue
\mciteSetBstMidEndSepPunct{\mcitedefaultmidpunct}
{\mcitedefaultendpunct}{\mcitedefaultseppunct}\relax
\EndOfBibitem
\bibitem[Chaimovich and Shell(2011)Chaimovich, and Shell]{Chaimovich2011}
Chaimovich,~A.; Shell,~M.~S. Coarse-graining errors and numerical optimization
  using a relative entropy framework. \emph{J. Chem. Phys.} \textbf{2011},
  \emph{134}, 094112\relax
\mciteBstWouldAddEndPuncttrue
\mciteSetBstMidEndSepPunct{\mcitedefaultmidpunct}
{\mcitedefaultendpunct}{\mcitedefaultseppunct}\relax
\EndOfBibitem
\bibitem[Schneider and Thiel(1989)Schneider, and Thiel]{Schneider1989}
Schneider,~W.; Thiel,~W. Anharmonic force fields from analytic second
  derivatives: Method and application to methyl bromide. \emph{Chem. Phys.
  Lett.} \textbf{1989}, \emph{157}, 367--373\relax
\mciteBstWouldAddEndPuncttrue
\mciteSetBstMidEndSepPunct{\mcitedefaultmidpunct}
{\mcitedefaultendpunct}{\mcitedefaultseppunct}\relax
\EndOfBibitem
\bibitem[Barducci \latin{et~al.}(2011)Barducci, Bonomi, and
  Parrinello]{Barducci2011}
Barducci,~A.; Bonomi,~M.; Parrinello,~M. Metadynamics. \emph{Wiley Interdiscip.
  Rev.: Comput. Mol. Sci.} \textbf{2011}, \emph{1}, 826--843\relax
\mciteBstWouldAddEndPuncttrue
\mciteSetBstMidEndSepPunct{\mcitedefaultmidpunct}
{\mcitedefaultendpunct}{\mcitedefaultseppunct}\relax
\EndOfBibitem
\bibitem[Chaimovich and Shell(2010)Chaimovich, and Shell]{Chaimovich2010}
Chaimovich,~A.; Shell,~M.~S. Relative entropy as a universal metric for
  multiscale errors. \emph{Physical Review E} \textbf{2010}, \emph{81},
  060104\relax
\mciteBstWouldAddEndPuncttrue
\mciteSetBstMidEndSepPunct{\mcitedefaultmidpunct}
{\mcitedefaultendpunct}{\mcitedefaultseppunct}\relax
\EndOfBibitem
\bibitem[Toxvaerd(1994)]{Toxvaerd1994}
Toxvaerd,~S. Hamiltonians for discrete dynamics. \emph{Phys. Rev. E}
  \textbf{1994}, \emph{50}, 2271\relax
\mciteBstWouldAddEndPuncttrue
\mciteSetBstMidEndSepPunct{\mcitedefaultmidpunct}
{\mcitedefaultendpunct}{\mcitedefaultseppunct}\relax
\EndOfBibitem
\bibitem[Toxvaerd(2013)]{Toxvaerd2013}
Toxvaerd,~S. Ensemble simulations with discrete classical dynamics. \emph{J.
  Chem. Phys.} \textbf{2013}, \emph{139}, 224106\relax
\mciteBstWouldAddEndPuncttrue
\mciteSetBstMidEndSepPunct{\mcitedefaultmidpunct}
{\mcitedefaultendpunct}{\mcitedefaultseppunct}\relax
\EndOfBibitem
\bibitem[Zhang \latin{et~al.}(Aug. 26--28, 2020)Zhang, Cooper, and
  De~Sa]{Zhang2020}
Zhang,~R.; Cooper,~A.~F.; De~Sa,~C. AMAGOLD: Amortized Metropolis adjustment
  for efficient stochastic gradient MCMC. International Conference on
  Artificial Intelligence and Statistics. Online, Aug. 26--28, 2020; pp
  2142--2152\relax
\mciteBstWouldAddEndPuncttrue
\mciteSetBstMidEndSepPunct{\mcitedefaultmidpunct}
{\mcitedefaultendpunct}{\mcitedefaultseppunct}\relax
\EndOfBibitem
\bibitem[Deng \latin{et~al.}(Jul. 13--18, 2020)Deng, Feng, Gao, Liang, and
  Lin]{Deng2020}
Deng,~W.; Feng,~Q.; Gao,~L.; Liang,~F.; Lin,~G. Non-convex learning via replica
  exchange stochastic gradient mcmc. Proceedings of the 37th International
  Conference on Machine Learning. Online, Jul. 13--18, 2020; pp
  2474--2483\relax
\mciteBstWouldAddEndPuncttrue
\mciteSetBstMidEndSepPunct{\mcitedefaultmidpunct}
{\mcitedefaultendpunct}{\mcitedefaultseppunct}\relax
\EndOfBibitem
\bibitem[Gorham and Mackey(Dec. 7--12, 2015)Gorham, and Mackey]{Gorham2015}
Gorham,~J.; Mackey,~L. Measuring sample quality with Stein's method. Advances
  in Neural Information Processing Systems. Montreal, QC, Canada, Dec. 7--12,
  2015\relax
\mciteBstWouldAddEndPuncttrue
\mciteSetBstMidEndSepPunct{\mcitedefaultmidpunct}
{\mcitedefaultendpunct}{\mcitedefaultseppunct}\relax
\EndOfBibitem
\bibitem[Garriga-Alonso and Fortuin(Jan. -- Feb., 2021)Garriga-Alonso, and
  Fortuin]{Garriga2020}
Garriga-Alonso,~A.; Fortuin,~V. Exact Langevin Dynamics with Stochastic
  Gradients. 3rd Symposium on Advances in Approximate Bayesian Inference.
  Online, Jan. -- Feb., 2021\relax
\mciteBstWouldAddEndPuncttrue
\mciteSetBstMidEndSepPunct{\mcitedefaultmidpunct}
{\mcitedefaultendpunct}{\mcitedefaultseppunct}\relax
\EndOfBibitem
\bibitem[Zhang \latin{et~al.}(May 6--9, 2019)Zhang, Li, Zhang, Chen, and
  Wilson]{Zhang2019cyc}
Zhang,~R.; Li,~C.; Zhang,~J.; Chen,~C.; Wilson,~A.~G. Cyclical Stochastic
  Gradient MCMC for Bayesian Deep Learning. 7th International Conference on
  Learning Representations. New Orleans, LA, USA, May 6--9, 2019\relax
\mciteBstWouldAddEndPuncttrue
\mciteSetBstMidEndSepPunct{\mcitedefaultmidpunct}
{\mcitedefaultendpunct}{\mcitedefaultseppunct}\relax
\EndOfBibitem
\bibitem[Heek and Kalchbrenner(Apr. 26--30, 2020)Heek, and
  Kalchbrenner]{Heek2020}
Heek,~J.; Kalchbrenner,~N. Bayesian Inference for Large Scale Image
  Classification. 8th International Conference on Learning Representations.
  Addis Ababa, Ethiopia, Apr. 26--30, 2020\relax
\mciteBstWouldAddEndPuncttrue
\mciteSetBstMidEndSepPunct{\mcitedefaultmidpunct}
{\mcitedefaultendpunct}{\mcitedefaultseppunct}\relax
\EndOfBibitem
\bibitem[Korattikara~Balan \latin{et~al.}(Dec. 7--12, 2015)Korattikara~Balan,
  Rathod, Murphy, and Welling]{Korattikara2015}
Korattikara~Balan,~A.; Rathod,~V.; Murphy,~K.~P.; Welling,~M. Bayesian dark
  knowledge. Advances in Neural Information Processing Systems. Montreal, QC,
  Canada, Dec. 7--12, 2015\relax
\mciteBstWouldAddEndPuncttrue
\mciteSetBstMidEndSepPunct{\mcitedefaultmidpunct}
{\mcitedefaultendpunct}{\mcitedefaultseppunct}\relax
\EndOfBibitem
\bibitem[Wang and Yoon(2021)Wang, and Yoon]{Wang2021}
Wang,~L.; Yoon,~K.-J. Knowledge Distillation and Student-Teacher Learning for
  Visual Intelligence: A Review and New Outlooks. \emph{IEEE Trans. Pattern
  Anal. Mach. Intell.} \textbf{2021}, \emph{44}, 3048--3068\relax
\mciteBstWouldAddEndPuncttrue
\mciteSetBstMidEndSepPunct{\mcitedefaultmidpunct}
{\mcitedefaultendpunct}{\mcitedefaultseppunct}\relax
\EndOfBibitem
\end{mcitethebibliography}


\end{document}